\documentclass[reprint,amsfonts, amssymb, amsmath,  showkeys,prx, superscriptaddress, twocolumn,longbibliography,nofootinbib]{revtex4-2}
\usepackage{float}
\makeatletter
\let\newfloat\newfloat@ltx
\makeatother
\usepackage[english]{babel}
\usepackage[utf8]{inputenc}
\usepackage{graphics}
\usepackage{selinput}
\usepackage[normalem]{ulem}
\usepackage[shortlabels]{enumitem}

\usepackage{braket}
\usepackage{amsthm}
\usepackage{mathtools}
\usepackage{physics}
\usepackage{xcolor}
\usepackage{graphicx}
\usepackage[left=16mm,right=16mm,top=35mm,columnsep=15pt]{geometry} 
\usepackage{adjustbox}
\usepackage{placeins}
\usepackage[T1]{fontenc}
\usepackage{lipsum}
\usepackage{csquotes}
\usepackage{bm}
\usepackage{ragged2e} 

\usepackage[linesnumbered,ruled,vlined]{algorithm2e}
\SetKwInput{kwInit}{Init}

\def\HC{\mathcal{H}}

\def\LC{\mathcal{L}}

\def\ad{^{\dagger}}

\newcommand{\fsnull}[1]{}
\newcommand{\old}[1]{}

\usepackage[makeroom]{cancel}
\usepackage[toc,page]{appendix}
\usepackage[colorlinks=true,citecolor=blue,linkcolor=magenta]{hyperref}

\usepackage{tikz}
\tikzset{every picture/.style=remember picture}

\usepackage[utf8]{inputenc}
\usepackage{graphicx}
\usepackage{xcolor}
\usepackage{amsmath}
\usepackage{amsthm}
\usepackage{bm}
\usepackage{bbm}
\usepackage{comment}
\usepackage{appendix}
\usepackage{mathdots}
\usepackage{lipsum}
\usepackage{verbatim}
\usepackage{natbib}
\usepackage{nccmath}
\usepackage{amsfonts} 

\newcommand{\dya}[1]{\ket{#1}\!\bra{#1}}

\newcommand{\AC}{\mathcal{A}}
\newcommand{\BC}{\mathcal{B}}
\newcommand{\CC}{\mathcal{C}}

\newcommand{\EC}{\mathcal{E}}

\newcommand{\OC}{\mathcal{O}}

\newcommand{\RC}{\mathcal{R}}
\newcommand{\SC}{\mathcal{S}}

\newcommand{\XC}{\mathcal{X}}

\renewcommand{\geq}{\geqslant}
\renewcommand{\leq}{\leqslant}

\DeclareMathOperator*{\argmin}{arg\,min}
\renewcommand{\vec}[1]{\boldsymbol{#1}}  

\newcommand*{\id}{\openone}

\newcommand{\bs}{\textsf{BS}}

\newcommand{\thv}{\vec{\theta}}

\def\be{\begin{equation}}
\def\ee{\end{equation}}
\def\bs{\begin{split}}
\def\e{\end{split}}
\def\ba{\begin{eqnarray}}
\def\bea{\begin{eqnarray}}

\def\tea{\end{eqnarray}}
\def\ea{\end{eqnarray}}
\def\eea{\end{eqnarray}}

\newtheorem{theorem}{Theorem}
\newtheorem{lemma}{Lemma}

\newtheorem{result}{Result}

\newtheorem{definition}{Definition}

\usepackage{amssymb}
\usepackage{dsfont}

\def\be{\begin{equation}}
\def\te{\end{equation}}
\def\ee{\end{equation}}
\def\ba{\begin{eqnarray}}
\def\bea{\begin{eqnarray}}

\def\tea{\end{eqnarray}}
\def\ea{\end{eqnarray}}
\def\eea{\end{eqnarray}}

\begin{document}

\title{Quantum Convolutional Neural Networks are Effectively Classically Simulable}

\author{Pablo Bermejo}
\affiliation{Information Sciences, Los Alamos National Laboratory, Los Alamos, NM 87545, USA}
\affiliation{Donostia International Physics Center, Paseo Manuel de Lardizabal 4, E-20018 San Sebasti\'an, Spain}
\affiliation{Department of Applied Physics, Gipuzkoa School of Engineering, University of the Basque
Country (UPV/EHU), Plaza Europa 1, 20018 San Sebastián, Spain}

\author{Paolo Braccia}
\affiliation{Theoretical Division, Los Alamos National Laboratory, Los Alamos, NM 87545, USA}

\author{Manuel S. Rudolph}
\affiliation{Ecole Polytechnique Fédérale de Lausanne (EPFL),  Lausanne CH-1015, Switzerland}

\author{Zo\"e Holmes}
\affiliation{Ecole Polytechnique Fédérale de Lausanne (EPFL),  Lausanne CH-1015, Switzerland}

\author{Lukasz Cincio}
\affiliation{Theoretical Division, Los Alamos National Laboratory, Los Alamos, NM 87545, USA}

\author{M. Cerezo}
\thanks{cerezo@lanl.gov}
\affiliation{Information Sciences, Los Alamos National Laboratory, Los Alamos, NM 87545, USA}

\begin{abstract}
    Quantum Convolutional Neural Networks (QCNNs) are widely regarded as a promising model for Quantum Machine Learning (QML). In this work, we analyze the most widely used variants of these models (i.e.,  tracing out- and measurement-based QCNNs), and we relate their heuristic success to two facts. First, that when randomly initialized, they can only operate on the information encoded in low-bodyness measurements of their input states. And second, that they are commonly benchmarked on ``locally-easy'' datasets whose states are precisely classifiable by the information encoded in these low-bodyness observables subspace. From these insights, we  argue that the QCNN's action on this subspace should be efficiently classically simulable. Indeed, we construct and train a purely classical QCNN surrogate—based on low-bodyness Pauli propagation, tensor networks, and classical shadows—that matches or outperforms standard QCNNs on all benchmark datasets and on up-to $1024$ qubits, thereby empirically realizing  our simulability claims. Our results can then be understood as highlighting a deeper symptom of QML: Models could only be showing heuristic success because they are benchmarked on simple problems, for which their action can be classically simulated. This insight points to the fact that non-trivial datasets are a truly necessary ingredient for moving forward with QML. To finish,  we discuss how our results can be extrapolated to classically simulate other architectures. 
\end{abstract}

\maketitle

\section{Introduction}

Quantum Machine Learning~\cite{chang2025primer} (QML) harnessed significant attention by riding on the combined heuristic success of classical machine learning~\cite{goodfellow2016deep,lecun2015deep,schmidhuber2015deep,bronstein2021geometric} and the computational promises of quantum computing~\cite{biamonte2017quantum,schuld2018supervised,cerezo2020variationalreview,bharti2021noisy,cerezo2022challenges,endo2021hybrid}. In the last few years, however, it has been recognized that QML models are prone to encountering serious trainability barriers--such as barren plateaus and local minima--that prevent their large-scale implementations~\cite{larocca2024review,mcclean2018barren,cerezo2020cost,cerezo2020impact,arrasmith2021equivalence,ragone2023unified,holmes2021connecting,bittel2021training,fontana2022nontrivial,anschuetz2022beyond,anschuetz2021critical,larocca2021diagnosing,bermejo2024improving,gil2024relation}. While several methods have been proposed to mitigate these issues~\cite{cerezo2020cost,pesah2020absence,khatri2019quantum,zhao2021analyzing,liu2021presence,miao2023isometric,larocca2021diagnosing,monbroussou2023trainability,cherrat2023quantum,fontana2023theadjoint,ragone2023unified,diaz2023showcasing,west2024provably,zhang2022escaping,park2023hamiltonian, wang2023trainability,larocca2022group,meyer2022exploiting,skolik2022equivariant,ragone2022representation,nguyen2022atheory,schatzki2022theoretical,kieferova2021quantum,rudolph2023trainability,letcher2023tight,rudolph2022synergy}, the recent work of Ref.~\cite{cerezo2023does} presents the intriguing connection between absence of barren plateaus and classical simulability. In particular, Ref.~\cite{cerezo2023does} hints at the fact that the very features that lead to provable absence of exponentially vanishing gradients, also serve as blueprints for classical algorithms that can simulate the action of the model, provided that one is given access to an initial data acquisition phase on a quantum computer.

Here it is important to highlight that the absence of barren plateaus is generally proved as an average statement (i.e., one shows that \textit{most} of the landscape does not exhibit a barren plateau). Correspondingly, certain barren plateau-free models are simulable on average (i.e., we can classically estimate the loss function for \textit{most} randomly sampled parameter settings). However, one could argue that the ability to classically simulate  the model at random points of the landscape is not practically useful, as the sections of the landscape that encode the solution to a problem are atypical. The question thus becomes: ``\textit{Can we not only simulate randomly sampled points of a barren plateau-free model, but also replicate a successful training over a simulated landscape?}'' Proving such a strong result becomes a much harder task, as one has to approach the problem in a  model- and even task-dependent way. 

In this work, we analyze the previous question by focusing on Quantum Convolutional Neural Networks (QCNNs)~\cite{cong2019quantum}.  QCNNs, either based on pooling layers where qubits are traced-out or measured, have shown heuristic success for supervised classification tasks based on classical~\cite{hur2021quantum, oh2020tutorial, baek2022scalable, gong2024quantum, kim2023classical, li2022image, bokhan2022multiclass, ovalle2023quantum, Chang2023Approximately, fan2023hybrid, li2020quantum, zeng2022multi, sebastianelli2021circuit, chalumuri2022quantum, zhang2019polsar, matsumoto2022full, chang2022quantum, aldoski2023impact, li2020quantum, fan2023hybrid, khan2023lightweight, chen2022quantumconvolutional, chen2021hybrid, delgado2022quantum,li2024quantum} and quantum data~\cite{monaco2022quantum, cea2024exploring, ferreira-martins2023detecting, caro2021generalization, gil-fuster2024understanding, nguyen2022atheory, cong2019quantum, maccormack2020branching, herrmann2022realizing, wrobel2022detecting, liu2023model,nguyen2022atheory,zapletal2024error}, and currently regarded as some of the promising QML architectures~\cite{acampora2025quantum,doosti2024brief,guan2021quantum,gupta2025systematic}. Indeed, these model have been shown to be barren plateaus-free in Ref.~\cite{pesah2020absence} for low entangled datasets. However,  while Ref.~\cite{cerezo2023does} gives some intuition as to why QCNNs should be classically simulable \textit{on average} (with an explicit algorithm recently presented in our companion manuscript of Ref.~\cite{angrisani2024classically}\footnote{That is, the results of Ref.~\cite{angrisani2024classically} imply that randomly initialized QCNNs are classically simulable, but this does not imply that the simulation will either be efficient or faithful throughout the whole training process. }), there is no rigorous and systematic study of their simulability across \textit{all} the relevant parts of the landscape which are accessed during training. 

Here we argue that the action of both tracing out- and measurement-based QCNNs is simulable, not just on average but in the stronger sense of enabling successful training, by  classical algorithms enhanced with classical shadows~\cite{cerezo2023does}. Indeed, we do not merely conjecture that such a classical surrogate could work—we explicitly construct it, train it on standard classical and quantum datasets (up to $1024$ qubits), and observe comparable or improved classification accuracies compared to standard QCNNs. Importantly, the fact that such classical simulation requires classical shadows means that a quantum computer might still be needed (to perform measurements  of the input data), but our construction fully avoids the need to  train an actual QCNN architecture on a quantum computer with a hybrid quantum-classical loop. This substantially reduces both the depths of circuits and number of circuits that need to be run on the quantum device. In fact, a universal quantum computer may no longer even be needed if the  quantum data can be obtained from an analogue quantum simulator or a more conventional quantum experiment. Importantly,  we will see that our results also suggest that likely no quantum computer is needed at all in the case of classical data.

\begin{figure}[t]
    \centering
    \includegraphics[width=1\linewidth]{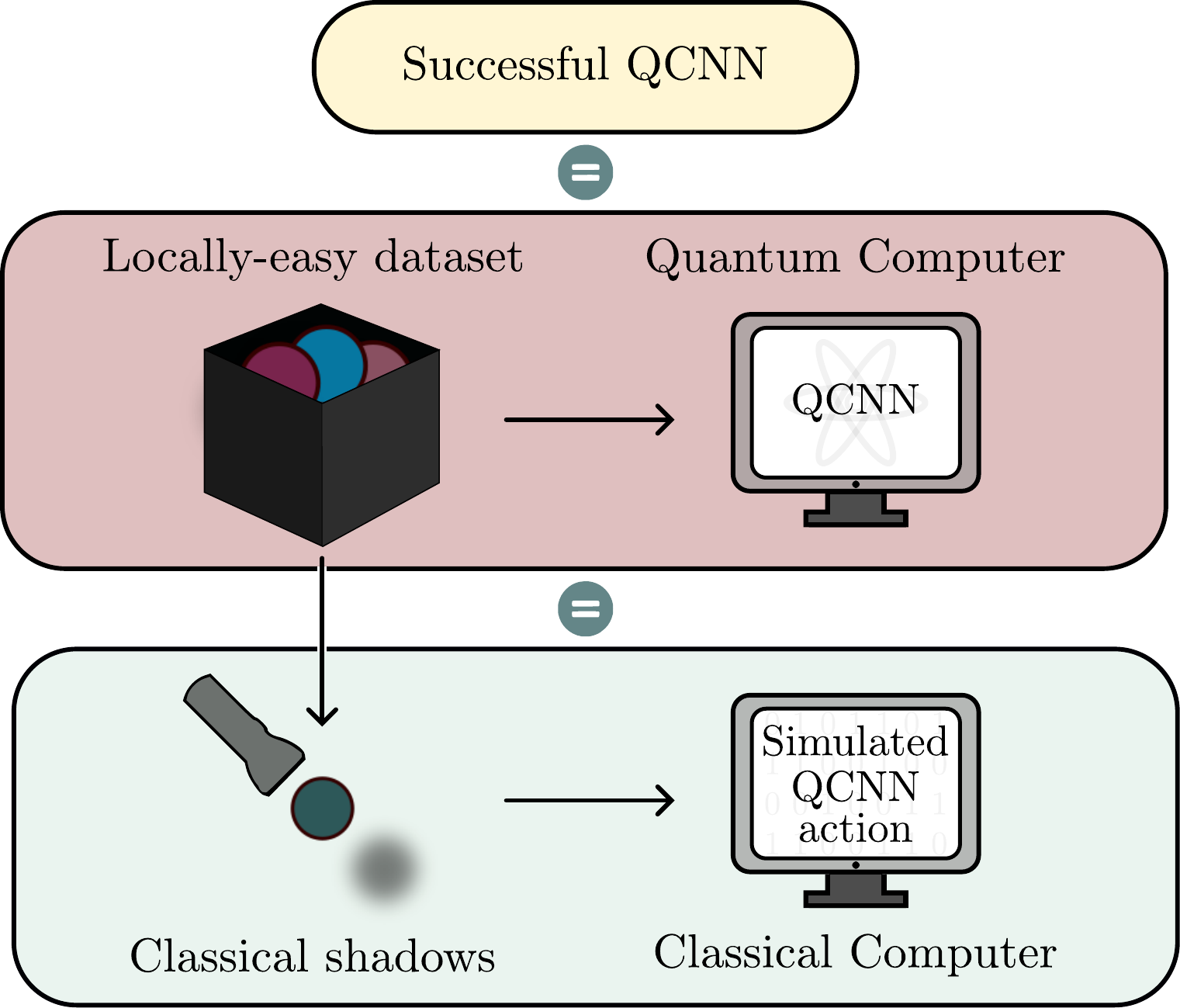}
    \caption{\textbf{Schematic representation of our main results.} We conceptualize the success of QCNN as a consequence of two facts: (1) When randomly initialized, they operate on a polynomially-sized subspace of low-bodyness observables, (2) They are benchmarked on locally-easy datasets that are  classifiable via the information encoded in low-bodyness measurements. The combination of these two facts allows us to show that there exist efficient classical algorithms that can simulate the action of the QCNN in this small subspace, provided that we are given access to Pauli classical shadows on the input data.    }
    \label{fig:schematic}
\end{figure}

To reach this conclusion, we will show the following results (schematically shown in Fig.~\ref{fig:schematic}). First, that randomly initialized tracing out- and measurement-based QCNNs  avoid barren plateaus by only extracting and processing the information encoded in low-bodyness (i.e., low-weight) observables of their input states. Second, we will show that the datasets (classical and quantum) used in the literature to demonstrate the power of QCNNs are ``\textit{locally-easy}''. That is, that they can be classified by a QCNN that is randomly initialized and trained on the information encoded in this polynomially-sized low-bodyness operator subspace.  
Then,  we show that the action of the QCNN can be replaced by a classically surrogate via a new variant of the LOWESA algorithm~\cite{fontana2023classical, rudolph2023classical} or by  tensor networks~\cite{orus2014practical}; provided that we are given access to local Pauli classical shadows on the dataset's states~\cite{huang2020predicting,elben2022randomized,sauvage2024classical}.   Crucially, these insights for quantum data also point to the ever-increasing realization that classical algorithms equipped with measurements from quantum computers appear to be an extremely promising path forward for quantum machine learning~\cite{huang2021power,cerezo2023does,huang2024learning}. 

While our results showcase that the heuristic success of QCNNs appears to be tied to locally-easy  datasets, this does not mean that examples could exist where the QCNN needs to leave the classically simulable regime to solve a given task. In fact, one can readily construct non-classically simulable datasets and QCNNs by embedding Shor's algorithm or  some cryptographic assumption into the problem (see for instance Refs.~\cite{cerezo2023does,jerbi2023shadows,gil2024relation}), but these are usually mathematical artifacts that are nowhere near related to realistic practical problems. Thus, the field of QML is in dire need of interesting, non-trivial datasets, in which models such as QCNNs can work without simultaneously being classically simulable. We present this as a challenge to the community and argue that until the viability of tracing out- and measurement-based QCNNs for such datasets can be demonstrated there is no reason to believe QCNNs will be useful. 

\section{Framework}

In this section  we briefly introduce the considered supervised classification QML setting, as well as recall the basic ingredients of a  QCNN, as defined in Ref.~\cite{cong2019quantum}. Then, we will present different notions of classical simulability, and discuss which one will be the focus of this work.

\subsection{Supervised learning with QCNNs}

To begin, let us recall that a supervised QML task is defined in terms of a data space $\mathcal{R}$, which belongs to the space of quantum states $\RC\subseteq\SC(\HC)$ of an $n$-qubit Hilbert space $\HC$, an integer-valued label space $\mathcal{Y}$, and an unknown function $f : \mathcal{R} \mapsto \mathcal{Y}$ that assigns a label $y_i = f(\rho_i)$ to the states in $\mathcal{R}$.  As such, our goal is to optimize a parametrized function $h_{\vec{\theta}}: \mathcal{R} \mapsto \mathcal{Y}$ to approximate the labels produced by $f$. The training of $h_{\vec{\theta}}$ is carried out by having repeated access to a dataset $T = \{(\rho_i, y_i)\}_{i}$, where $\rho_i$ is drawn from $\mathcal{R}$ according to some probability distribution, and $y_i \in \mathcal{Y}$ are the associated labels. In this work we will consider two scenarios of interest corresponding to the classification of classical and quantum data~\cite{cerezo2022challenges}. For the case of classical data,  we assume that the states in  $\mathcal{R}$ were produced by encoding some real valued classical vector $x_i\in\XC$  into a quantum state by means of an encoding map $\EC:\XC\rightarrow \RC$. Then, for  quantum data, the states in $\RC$ are produced by  some quantum mechanical process of interest.

\begin{figure}[t]
    \centering
    \includegraphics[width=1\linewidth]{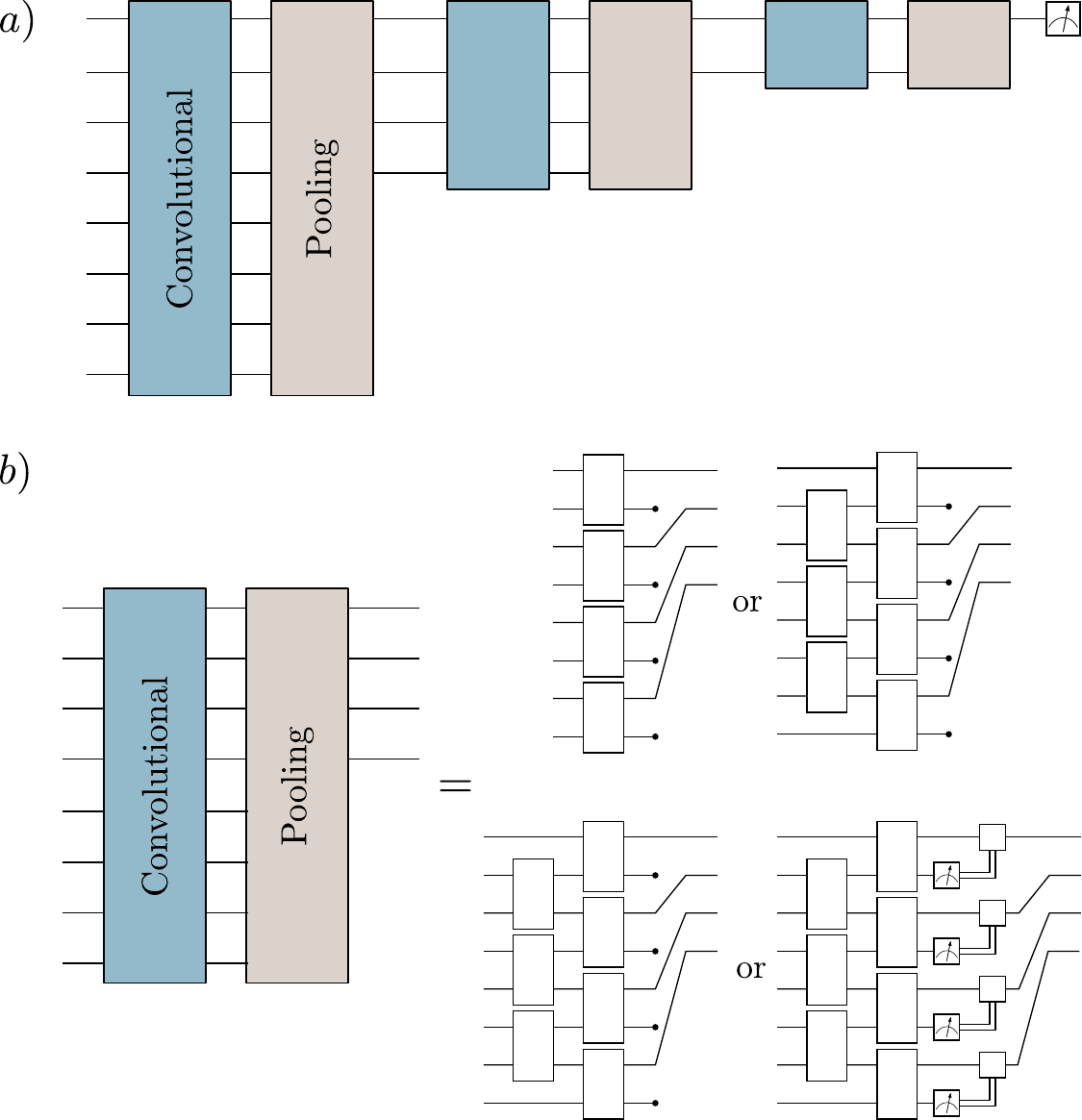}
    \caption{\textbf{QCNN architecture.} (a) A QCNN is composed of alternating convolutional and pooling layers. In the convolutional layers, information is usually being processed by parametrized quantum gates. In the pooling layers, the dimension of the QCNN feature space is reduced by tracing out or measuring qubits. By design, QCNNs have a depth that only scales logarithmically with the number of qubits $n$, and the measurements at its output are local. (b) Examples of tracing out- and measurement-based QCNNs. Convolutional layers are composed of two-local gates acting on nearest neighbors. In a tracing out QCNN half of the qubits are being traced out during the pooling layer; whereas in a measurement-based QCNN, half of the qubits are measured and the outcomes control unitaries on one of their nearest neighbors.  }
    \label{fig:QCNN}
\end{figure}

In this paper we will focus on QML schemes based on QCNN architectures~\cite{cong2019quantum}. As shown in Fig.~\ref{fig:QCNN}(a), a QCNN is composed of a sequence of convolutional and pooling layers,  with the first being used to coarse-grain the information of the input quantum-states, and the latter aimed at reducing the QCNN's feature space by tracing out or measuring qubits. Following the definition in Ref.~\cite{cong2019quantum}, we therefore can define a generic QCNN as the composition of quantum channels of the form
\begin{equation}\label{eq:qcnn-map}
    \Phi_{\vec{\theta},\vec{\lambda}} =  \bigcirc_{l=1}^{L} \left(P_{l}^{\vec{\lambda}_{l}} \circ C_{l}^{\vec{\theta}_{l}}\right),
\end{equation}
where $C$ and $P$ are parameterized by $\vec{\theta}$ and $\vec{\lambda}$, respectively, representing convolutional and pooling quantum maps. Convolutional maps $C_l^{\vec{\theta}_{j}}:\SC(\HC_l)\rightarrow\SC(\HC_l)$ preserve the size of the quantum registers they act on, while pooling maps reduce it, i.e., $P_l^{\vec{\lambda}_{l}}:\SC(\HC_l)\rightarrow\SC(\HC_{l+1})$ such that $\dim(\HC_{l+1})< \dim(\HC_{l}) $. Typically, in a pooling layer half of the qubits are discarded or measured, meaning that the total number of layers satisfies $L\in\Theta(\log(n))$, i.e., for typical QCNN architectures the number of layers scales logarithmically with the system size.

While Eq.~\eqref{eq:qcnn-map} allows for a wide variety of QCNNs, in this work we consider their most common instantiations: (1) tracing out based QCNNs, where one uses unitary convolution maps and simply traces-out of qubits (according to some pattern) in the  pooling layers, and (2) measurement-based QCNNs, where qubits are measured in the pooling layers, and their results are used to control unitaries in neighboring sites. To begin, we will mostly focus on tracing out based QCNNs. As we will see below in our numerical results section, this choice is motivated from the fact that these simpler architectures can be found to achieve similar, or even better, success than other more general ones (e.g., measurement-based QCNNs), indicating that additional levels of complexity might not be needed.  Hence, in what follows we will assume that 
\begin{equation}\label{eq:standard_qcnn_conv_pool}
\begin{aligned}
        C_l^{\vec{\theta}_l} (\cdot) &= U_l(\vec{\theta}_l) (\cdot) U_l^\dagger(\vec{\theta}_l), \\
        P_l^{\vec{\lambda}_{l}}(\cdot) &= \Tr_{t_l}[\, \cdot \,],
\end{aligned}
\end{equation}
where $t_l$ determines the set of qubits being traced-out at the $l$-th layer. Given that pooling layers are not parametrized we will denote the tracing out QCNN map as $ \Phi_{\vec{\theta}}$.

At the output of this quantum neural network, we measure  an observable $O$ acting on $\OC(1)$ qubits to compute the predicted label
\begin{equation}\label{eq:predicted-label}
    \Bar{y}_i(\thv) = \Tr[\Phi_{\vec{\theta}}(\rho_i)O]\,.
\end{equation}
These predictions are then used to evaluate a loss function $\LC_{\thv}(\{\Bar{y}_i(\thv)\}_i,\{y_i\}_i)$ that quantifies how close the predicted labels are from the true ones on the training dataset $T$. For instance, one can use the mean-squared error or cross-entropy loss functions that are widely used in classical machine learning. Then, the QCNN's parameters are trained by leveraging the power of classical optimizers that solve the optimization task
\begin{equation}\label{eq:optimization-task}
    \thv^*=\argmin_{\thv} \LC(\{\Bar{y}_i(\thv)\}_i,\{y_i\}_i)\,.
\end{equation}
Once the optimal set of parameters $\thv^*$ are found, these are used to define the trained model $h_{\thv^*}$ (which is a function of the expectation values of Eq.~\eqref{eq:predicted-label})  to test its performance on new and previously unseen data points.

The success of the QML model in solving Eq.~\eqref{eq:optimization-task} hinges on several factors. For instance,  one needs to efficiently navigate the (highly non-convex) optimization landscape and reach its global minima. Here, it has been shown that several trainability barriers can exist in quantum neural network-based schemes such as the presence of barren plateaus (i.e., gradients that vanish exponentially with $n$) or sub-optimal local minima where the optimizer can get trapped in~\cite{larocca2024review,mcclean2018barren,cerezo2020cost,cerezo2020impact,arrasmith2021equivalence,ragone2023unified,holmes2021connecting,bittel2021training,fontana2022nontrivial,anschuetz2022beyond,anschuetz2021critical,larocca2021diagnosing,bermejo2024improving}. Fortunately, it has been shown that QCNNs do not exhibit barren plateaus~\cite{pesah2020absence}, and their heuristic success provides solid evidence that while local minima exist, they do not constitute a fundamental obstacle for QCNN trainability~\cite{hur2021quantum, oh2020tutorial, baek2022scalable, gong2024quantum, kim2023classical, li2022image, bokhan2022multiclass, ovalle2023quantum, Chang2023Approximately, fan2023hybrid, li2020quantum, zeng2022multi, sebastianelli2021circuit, chalumuri2022quantum, zhang2019polsar, matsumoto2022full, chang2022quantum, aldoski2023impact, li2020quantum, fan2023hybrid, khan2023lightweight, chen2022quantumconvolutional, chen2021hybrid, delgado2022quantum,monaco2022quantum, cea2024exploring, ferreira-martins2023detecting, caro2021generalization, gil-fuster2024understanding, nguyen2022atheory, cong2019quantum, maccormack2020branching, herrmann2022realizing, wrobel2022detecting, liu2023model}. 

While these two features on their own have made QCNNs one of the most promising QML models, there is one aspect of their performance that has not been discussed enough: \textit{ Whether they can be classically simulated, or not\footnote{It is clear that QCNNs with trivial input states such as the all zero-state are classically simulable via tensor networks. Here we instead wonder if their action is simulable even if the initial state does not admit a simple classical representation.  }.} Indeed, in a typical quantum neural network-based QML setting one assumes that the evaluation of the loss function $\LC(\{\Bar{y}_i(\thv)\}_i,\{y_i\}_i)$, or more precisely of the expectation values $\Bar{y}_i(\thv)$, requires the use of a quantum computer, as it is presumed that the information processing capabilities of the quantum model map are hard to simulate.  However, if a model is classically simulable, then  the implementation of its parameterized quantum circuit on a quantum computer is unnecessary.

\subsection{Notions of classical simulability}
When studying the classical simulability of a QML model, one usually focuses on whether there exists a classical algorithm $\AC$ that takes as input an efficient description of the problem (i.e., a gate sequence that prepares each $\rho_i$ from a fiduciary state, a sequence of operations in the circuit, and a representation of the measurement operator) and returns an estimation $\widehat{\LC}(\{\Bar{y}_i(\thv)\}_i,\{y_i\}_i)$ of the loss function~\cite{aaronson2004improved,jozsa2008matchgates,huang2023learning,huang2022learningmanybodyhamiltonians,goh2023lie,cerezo2023does}. In this framework, one generally focuses on the error  $|\LC(\{\Bar{y}_i(\thv)\}_i,\{y_i\}_i)-\widehat{\LC}(\{\Bar{y}_i(\thv)\}_i,\{y_i\}_i)|$ for a given set of parameters $\thv$, or on the average error $\mathbb{E}_{\thv}|\LC(\{\Bar{y}_i(\thv)\}_i,\{y_i\}_i)-\widehat{\LC}(\{\Bar{y}_i(\thv)\}_i,\{y_i\}_i)|$ over the whole landscape. The results in our companion work of Ref.~\cite{angrisani2024classically} can be used to provide bounds on the average error, showing that one can approximate the loss function to an arbitrarily small constant precision with polynomial time and sample complexity via low-bodyness (or low-Pauli-weight) approximations~\cite{huang2023learning}. 

While the results in Ref.~\cite{angrisani2024classically} show that random points in the QCNN's loss function landscape are easy to classically simulate with high probability, they do not imply that we can simulate the regions of the landscape that are relevant  during training. To account for this limitation, we instead focus on a more pragmatic definition of classical simulability where the ultimate goal is to solve  the task of interest~\cite{gil2024relation}. In particular, we will follow the setting of ``\textit{classical simulation enhanced with quantum experiments}'' (or more precisely, ``\textit{classical simulation enhanced with efficient shadow tomography}'') defined in Ref.~\cite{cerezo2023does}. Here, one is allowed access to a quantum computer for an initial data acquisition phase during which  one can prepare copies of the states in $\RC$, apply some operations, and independently measure them via some efficient tomographic  classical shadow
techniques~\cite{huang2020predicting,elben2022randomized,sauvage2024classical}. Once this initial data acquisition phase is over, one can no longer access the quantum computer. As such, we say that a model is classically simulable if there exists a classical algorithm $\AC$ that takes as input an efficient description of the problem, as well as the data obtained from quantum devices in an initial data acquisition phase, and returns an estimation $\widehat{\LC}(\{\Bar{y}_i(\thv)\}_i,\{y_i\}_i)$ such that the solution of
\begin{equation}\label{eq:optimization-task-2}
    \thv^*=\argmin_{\thv} \widehat{\LC}(\{\Bar{y}_i(\thv)\}_i,\{y_i\}_i)
\end{equation}
leads to a simulated trained model $\widehat{h}_{\thv^*_c}$ capable of achieving classification accuracies that are comparable to those of $h_{\thv}$. 

Note that this definition of classical simulability is intimately tied to the recently introduced notion of \textit{dequantization in QML} (Definition 3 in Ref.~\cite{gil2024relation}), with the addendum that we allow the classical algorithm access to measurements obtained from a quantum computer during an initial data acquisition phase.  As such, our approach to classical simulability can be regarded as a substantial dequantization of QML whereby the usage of the quantum device is reduced to just collecting shadows of the training data. While this is not a full dequantization in the sense that some form of quantum experiment may still be required, we stress that this experiment (state preparation followed by local Pauli measurements) is significantly simpler and so a full universal quantum computer may no longer be needed. As such, within the grander scheme of classical simulability of parametrized circuit-based QML models, we will say that if there is no need to run the parametrized quantum circuit on a quantum device, one can consider that the QML model's information processing abilities are dequantized.

\section{Conceptualizing the success of QCNN\protect\lowercase{s}}

In this section we present the basic ingredients to understand what makes QCNNs successful. Then, we will argue that these same features can be exploited to classically simulate QCNN-based QML models via classical algorithms enhanced with classical shadows.

\subsection{Randomly initialized QCNN\protect\lowercase{s}}

To begin, let us begin by considering a unitary tracing out QCNN architecture as in  Eq.~\eqref{eq:standard_qcnn_conv_pool}, where the convolutional layers are composed of general parametrized two-qubit gates acting on nearest neighboring qubits in a pattern such as those of  Fig.~\ref{fig:QCNN} (b, to). Then, let $\Phi_{\vec{\theta}}\ad(O)$ denote the Heisenberg-evolved measurement operator, and we will expand this operator in the Pauli basis and define the bodyness of any given Pauli as the number of qubits it acts non-trivially on.  In what follows, we will assume that when initializing the QCNN, we sample the parameters $\thv$ such that each local gate forms an independent local $2$-design over $\mathbb{U}(4)$~\cite{dankert2009exact,harrow2009random,brandao2016local,hunter2019unitary,haferkamp2022random,schuster2024random}. By means of the Weingarten calculus~\cite{collins2006integration,puchala2017symbolic,mele2023introduction} we can find that the following result holds (see the appendix for additional details):
\begin{result}[Informal]\label{res:bodyness} 
    Consider tracing out QCNNs where the convolutional layers are composed of general parametrized two-qubit gates acting on nearest neighboring qubits in a pattern such as those of  Fig.~\ref{fig:QCNN} (b, top left). In average, the contribution in $\Phi_{\vec{\theta}}\ad(O)$ of a given Pauli with bodyness $k$, decays exponentially with $k$.
\end{result}

A direct implication of Result~\ref{res:bodyness} is that when computing the inner product $\Tr[\rho \Phi_{\vec{\theta}}\ad(O)]$, the dominant contributions in the predicted labels of Eq.~\eqref{eq:predicted-label} will essentially only arise from $\OC(1)$-bodyness measurements over the initial state.  As such, we can see that, when randomly initialized, the QCNN will start its training by only processing the information encoded in the polynomially-sized subspace of low-bodyness observables in their input data. In fact, these realizations can be directly linked to the proof of absence of barren plateaus in QCNNs, as a careful revision of Ref.~\cite{pesah2020absence} reveals that the large components in the loss function's gradients arise precisely from the local terms in the Heisenberg-evolved measurement operator $\Phi_{\vec{\theta}}\ad(O)$. 

Importantly, we note that this conclusion is further strengthened by the results in our companion paper~\cite{angrisani2024classically}, where we show that, on average, the error in approximating the predicted labels via $\Phi_{\vec{\theta}}\ad(O)$ or via a low-$k$ body approximation of this operator (where one truncates all the contributions from Paulis with bodyness higher than $k$ after each layer, or after each gate), decays exponentially with $k$. Specifically, it decays as $\OC\left(\left(\frac{2}{3}\right)^k\right)$. Indeed, while this observation is used in Ref.~\cite{angrisani2024classically} to provide a classical algorithm that requires only polynomial time and sample complexity to simulate a QCNN on average, we note that the previous work does not perform end-to-end training on a simulated architecture. 

At this point we find it important to highlight the fact while we have shown that the dominant terms of $\Tr[\rho \Phi_{\vec{\theta}}\ad(O)]$ arise from local observables over the initial state, these can be exponentially suppressed if the  input data is too entangled (e.g., follows a volume law of entanglement~\cite{leone2022practical,thanaslip2021subtleties}). In this case, the dominant information that the QCNN operates on might not be encoded in this small subspace, and the theoretical guarantees for the model are no longer valid. Namely, it follows  from the results of Ref.~\cite{pesah2020absence} that QCNNs can exhibit barren plateaus when the data states are too entangled. 

Then, we note that while Result~\ref{res:bodyness} was theoretically derived for QCNNs where the convolutional layers are composed of general parametrized two-qubit gates acting on nearest neighboring qubits in a pattern such as those of  Fig.~\ref{fig:QCNN} (b), one can also generalize such result to other more architectures (e.g., where the convolutional layer has more gates) using the techniques in~\cite{braccia2024computing}.

To finish, we note that the results in Result~\ref{res:bodyness} are only an average statement, meaning that they only hold for randomly initialized QCNNs. Indeed, they do not provide a practical mean to simulate the action of the QCNN throughout training, when the parameters are not randomly sampled, but driven by an optimizer. Below we explicitly show how to construct a surrogate of the model and show that we can train on it to solve standard classification problems. Importantly, we note that our numerical simulations are a heuristic extension of the classical average case simulability entailed by Result~\ref{res:bodyness}, and should not be considered a bridge between average-case and worst-case scenarios. Indeed, we note below that that QCNN simulation will be successful only for specific datasets. 

\subsection{Locally-easy datasets}

In the previous sections we have argued that for random initializations the QCNN's training process is initially guided by the low-bodyness information over the input states, and that the loss function will exhibit large gradients if the initial states are not too entangled~\cite{pesah2020absence}. Hence,  we know that the QCNN will be able to -- at least -- take the first few training steps. From here, it is reasonable to expect that, if the landscape is sufficiently well-behaved  (e.g., local minima do not prevent training) and a good solution lies also within this subspace, then the  QCNN could  accurately solve the task at hand. Evidently,  analytically proving that all of these conditions are met could be an extremely hard task. However, we can define the following class of datasets for which QCNNs could perform well: 
\begin{definition}[Locally-easy dataset]\label{def:local-easy}
     Given a dataset for a supervised QML classification task, we will say that it is locally-easy if it can be classified using the expectation values of low-bodyness observables of the input quantum states. 
\end{definition}

Evidently, having a locally-easy dataset is a necessary condition for the optimization curve to lie within the low-bodyness subspace, but it is nonetheless not a sufficient one. It is entirely possible that there exist initial points that are only connected to the global minima through paths that need to leave the subspace. However, as we will see below, we compiled classical and quantum datasets used in the literature to  benchmark QCNNs, and we  heuristically show not only that they are all locally-easy, but the classification task can be solved by remaining in the low-bodyness subspace.   

To finish, we find it important to note that in the appendices we present results strengthening our claim that the considered datasets are locally-easy. Namely, we consider the XXX bond-alternating dataset (see below for more details),  generate $n=100$ qubits states, estimate expectation values on all local Pauli operators--plus a few two-qubit ones--and perform perfect classification using a simple random forest algorithm. These result show that local observables contain all relevant information for classification, as per Definition ~\ref{def:local-easy}.

\subsection{Measurement-based QCNNs}

To finish, let us now consider a measurement-based QCNN. For this purpose, we recall that the action of a local pooling operation on the $l$-th layer, where qubit $j$ is measured, and the outcome is used to control a trainable unitary $V(\vec{\lambda},x)$  on qubit $j'$ is given by 
\begin{equation}
P_{l,j}^{\vec{\lambda}_{l}}(\cdot)=\sum_{x_j\in\{0,1\}}p(x_j)\dya{x_j}\otimes V_{j'}(\vec{\lambda},x)(\, \cdot \,)\dya{x_j}\otimes V\ad_{j'}(\vec{\lambda},x)\,.\nonumber
\end{equation}
where $p(x)=\Tr[\dya{x} \rho_{l,j} ]$ and $\rho_{l,j}$ is the reduced state of the $j$-th qubit at the $l$-th layer. As such, the $l$-th pooing layer takes the form
\begin{equation}\label{eq:measurement_qcnn_conv_pool}
        P_l^{\vec{\lambda}_{l}}(\cdot) = \prod_{j\in t_l}P_{l,j}^{\vec{\lambda}_{l}}(\cdot).
\end{equation}
where we recall that $t_l$ determines the set of qubits being measured at the $l$-th layer. 

By using again the Weingarten calculus we  find that the following result holds (see the appendix for a proof):
\begin{result}[Informal]\label{res:bodyness-measurements} 
    Consider measurement-based QCNNs where the convolutional layers are composed of general parametrized two-qubit gates acting on nearest neighboring qubits in a pattern such as those of  Fig.~\ref{fig:QCNN} (b, bottom left). In average, the contribution in $\Phi_{\vec{\theta},\vec{\lambda}}\ad(O)$ of a given Pauli with bodyness $k$, decays exponentially with $k$.
\end{result}
Just as in tracing out-based QCNNs, Result~\ref{res:bodyness-measurements} shows that randomly initialized measurement-based QCNN also start their training by only processing the information encoded in the polynomially-sized subspace of low-bodyness observables in their input data.  This realization is related to the absence of barren plateaus in this architecture. In fact, we can also show that (independently of the convolutional layer architecture)
\begin{equation}\label{eq:conce}
   {\rm Var}_{\vec{\theta}}\left[\Tr\left[\Phi_{\vec{\theta}}(\rho)  O\right]\right]\leq {\rm Var}_{\vec{\theta},\vec{\lambda}}\left[\Tr\left[\Phi_{\vec{\theta},\vec{\lambda}}(\rho)  O\right]\right]\,,
\end{equation}
where $\Phi_{\vec{\theta}}$ denotes the tracing out QCNN channel. Equation.~\eqref{eq:conce} indicates that the variance of measurement-based QCNNs is larger than that of tracing out ones, showing  that adding measurements to the architecture decreases the model's  expressive power~\cite{sim2019expressibility,holmes2021connecting}.

\section{Numerical results for classically simulated QCNN\protect\lowercase{s}}\label{section:numerical results}

As argued above, QCNNs simulability arises from the fact that they are initialized, explore and end their training in the polynomially-large subspace of low-bodyness observables. We now use this fact to classically simulate QCNNs and show that its information processing capabilities can be emulated by classical computers for essentially all datasets used  in the literature. As such, the purpose of this section is three-fold:
\begin{enumerate}
    \item Proving that datasets commonly used  in the literature as heuristic evidence for QCNNs are locally-easy.  In particular, we will see that simple Pauli classical shadows are sufficient to extract the relevant features of the data. 
    \item Illustrating the classical simulation of QCNNs, as well as the scalability of the proposed techniques (we simulate and train QCNNs with upto 1024 qubits). 
    \item Studying how the performance of the simulated \mbox{QCNNs} depends on the amount of measurements used during the shadow tomographic procedure.  
\end{enumerate}

The end-to-end simulations of tracing out-based QCNNs presented in this section are based on  two techniques which are constructed to only process the information encoded in low-bodyness measurements of the input states. In particular, they accomplish this goal by truncating the Heisenberg-evolved measurement operators to bodyness $\OC(1)$, which allows us to guarantee an efficient representation of this backwards-evolved operator. As detailed in Appendix~\ref{apx:classical_methods}, the first simulation method used to classify quantum data is based on the LOWESA algorithm~\cite{fontana2023classical,rudolph2023classical}. Then, we classify classical data using a  tensor network technique with restricted bodyness. In addition, in Appendix~\ref{app:sec:meas-QCNN}, we show that tensor networks can also efficiently simulate measurement-based QCNNs for quantum data.   

We will split our numerical results into quantum datasets and classical datasets, and we employ different simulation methods for each of these families. This will showcase the diversity of classical methods that enable the simulation of the QCNN action, as well as their broad applicability. In all cases, we find that the simulated QCNN reaches similar, or better performances than those reported in the literature for ``full QCNNs'' (i.e., not restricted to the subspace of low-bodyness observables). More importantly, we show that our finite sample shadow-based simulations  can outperform full QCNNs trained with no finite sampling.

\subsection{Quantum datasets}

The problem of classifying quantum phases of matter is oftentimes considered as a key application of QML, since it inherently requires preparing quantum states of interest in a quantum device. However, despite the quantum nature of the data, we will argue that in the standard test cases used to demonstrate QCNNs the local information  encoded in low-bodyness measurements suffices   to classify the states.  We establish this by showing that we can train a simulated QCNN (using the LOWESA-based Pauli propagation methods outlined in the appendix) via Pauli classical shadows on the dataset states. Notably, the classical, weight-truncated QCNN is self-consistently trained to solve the classification tasks, not to faithfully emulate the training of an exact QCNN. 

In particular, we focus on four popular classification tasks based on the one-dimensional Heisenberg Bond-Alternating XXX model \cite{Kitazawa1996phase}, Haldane chain \cite{haldane1983nonlinear}, Axial next-nearest neighbor Ising (ANNNI) model \cite{elliott1961phenomenological}, and Cluster Hamiltonian \cite{suzuki1971relationship}. The datasets are composed of an evenly distributed number of ground-states per phase (obtained via  Density Matrix Renormalization Group (DMRG) from ITensors.jl software library \cite{fishman2022itensor} for the Julia programming language \cite{bezanson2017julia}). 
Then, the tomographic data is computed by performing standard Pauli classical shadows~\cite{huang2020predicting,elben2022randomized,sauvage2024classical}. In all cases,   we simulate the action of a QCNN as in Fig.~\ref{fig:QCNN}(b, right), where each block is a general two qubit unitary with $15$ parameters, and where we trace-out half of the remaining qubits in each pooling layer. 
The training of the simulated QCNN is performed by using a  cross-entropy based loss function and leveraging the LBFGS~\cite{liu1989limited} optimizer for the parameter update. 

Here we note that for the Bond-Alternating XXX model and the Haldane chain we will consider a binary classification task, and we will showcase the power of our simulations by considering spin systems of $n = 1024$ and $n = 512$ qubits, respectively. Then, for the  ANNNI model and the Cluster Hamiltonian, we will focus on multi-class classification and restrict ourselves to QCNNs acting on $n = 32$ qubits, so that we can compare our results to those previously obtained in the literature. 

\subsubsection{Heisenberg bond-alternating XXX model}

The Hamiltonian of the Heisenberg Bond-Alternating XXX model reads
\begin{equation}\label{eq:alt_xxx_model}
H = \sum_{i=1}^{n-1}J_i \left(X_iX_{i+1}+Y_iY_{i+1}+Z_iZ_{i+1}\right)\,,
\end{equation}
where $X_i$, $Y_i$ and $Z_i$ denote the Pauli operator acting on the $i$-th qubit, and where $J_i=J_1(J_2)\geq 0$ for $i={\rm even}({\rm odd})$. In the thermodynamic limit $n\rightarrow\infty$, the ground state of this model presents a phase transition controlled by the ratio $J_2/J_1$, such that $J_2<J_1$ corresponds to a trivial phase whereas when $J_2>J_1$ the ground-space is topologically protected. See Fig.~\ref{fig:XXX-diagram}(a) for a sketch of the phase diagram.

\begin{figure}[t]
    \centering
    \includegraphics[width=\columnwidth]{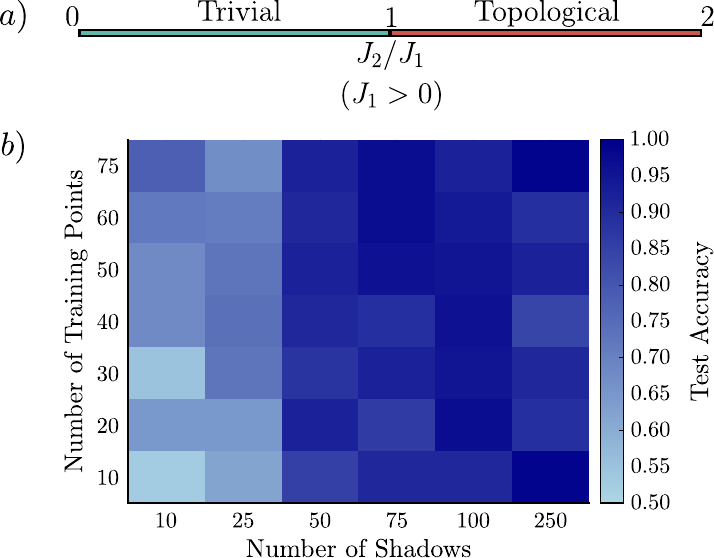}
    \caption{\textbf{Bond-Alternating XXX model.} a) Phase diagram for the Hamiltonian in Eq.~\eqref{eq:alt_xxx_model}. b)  Test classification accuracy for the simulated QCNN acting only on the low-bodyness operator subspace. We show the accuracy as a function of the number of training points and Pauli classical shadows on each state of the dataset.  }
    \label{fig:XXX-diagram}
\end{figure}

Here we consider a chain of $n = 1024$ qubits. To our knowledge, this constitutes the largest QCNN implementation with classical shadows (see also~\cite{zapletal2024error} for an implementation with large-scale matrix-product states), and we find it important to highlight that the simulation was performed with resources available on a modern laptop. At the output of a QCNN we measure the $Z_1$ operator. Throughout the simulation we cap the maximum bodyness of the Heisenberg-evolved measurement operator at two. We use a dataset of $100$ states and perform the training over $200$ iterations, averaged over $5$ different runs. To further strengthen our claim that just a small amount of information from the operator space is needed for the classification, we remark that only the first $400$ operators with the largest variance across the shadows dataset were used to train this model (see the appendix).  The classification results are shown in Fig.~\ref{fig:XXX-diagram}(b), where we see that by taking only $100$ classical shadows per data point one can consistently achieve test accuracies above $90\%$.  This last fact is extremely important as the total number of measurements shots used to train our classical QCNN never goes above $ 20000$ for the whole experiment. When compared against the resources needed to train a QCNN on a quantum computer, where at least $5000$--$10000$ shots are used to estimate the loss in each iteration step per data point, we can see that the amounts of quantum resources used in our scheme is order of magnitude smaller than standard QCNN training. Therefore, not only we can classically simulate the action of QCNNs, we can do so in an extremely measurement-frugal way.

\begin{figure}[t]
    \centering
\includegraphics[width=.9\columnwidth]{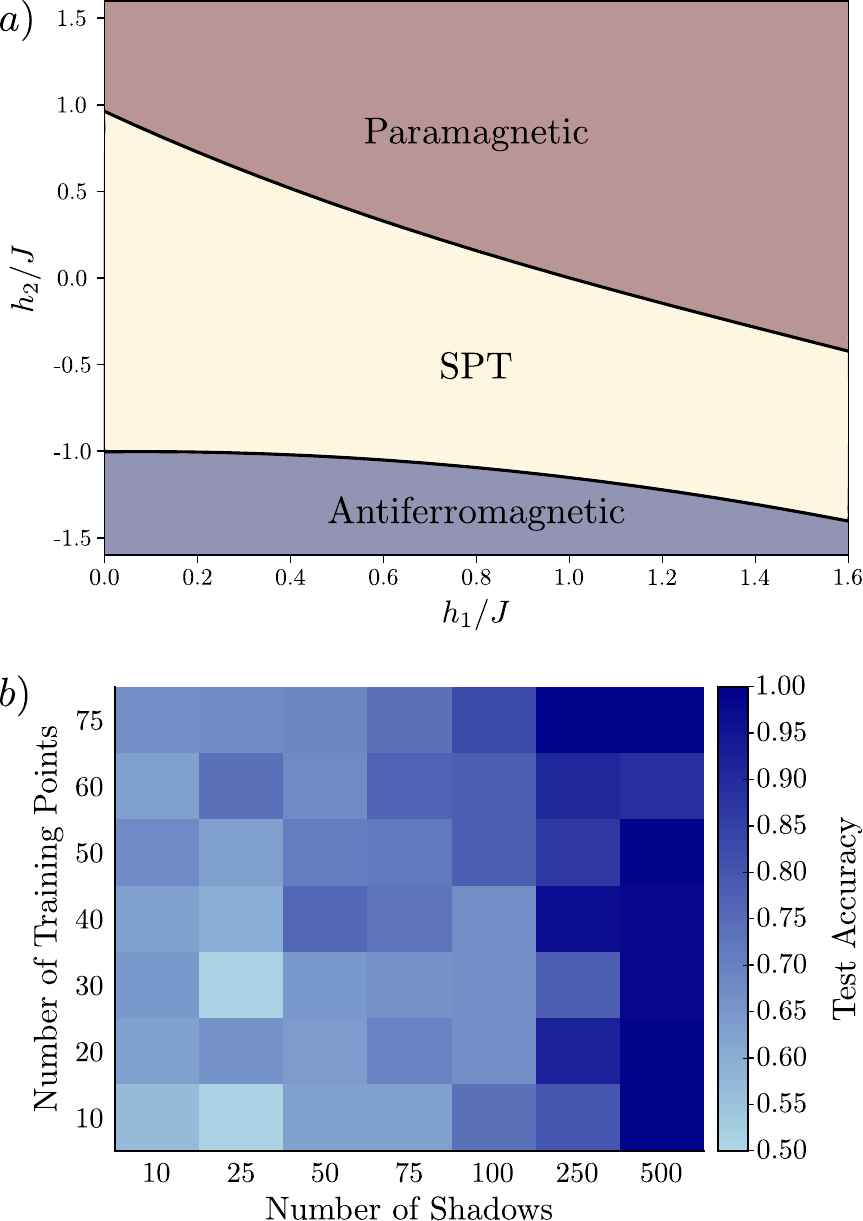}
    \caption{\textbf{Haldane Chain.} a) Phase diagram for the Hamiltonian in Eq.~\eqref{eq:haldane_model}. b)  Test classification accuracy for the simulated QCNN acting only on the low-bodyness operator subspace. We show the accuracy as a function of the number of training points and Pauli classical shadows on each state of the dataset.  }
    \label{fig:haldane-diagram}
\end{figure}

\begin{figure*}[t]
    \centering
    \includegraphics[width=\linewidth]{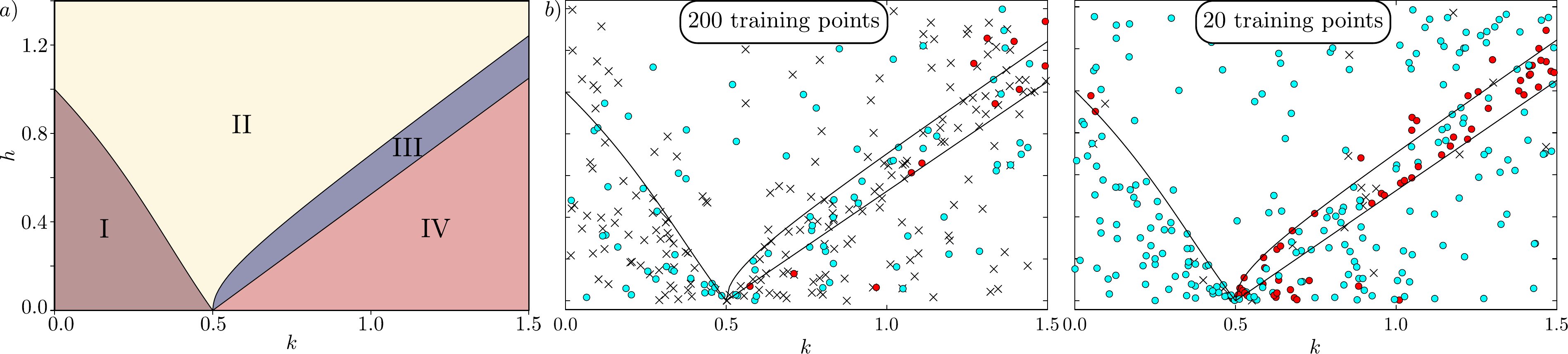}
    \caption{\textbf{ANNNI model.} a) Phase diagram for the Hamiltonian in Eq.~\eqref{eq:ANNNI_model}. The phases are: (I) ferromagnetic, (II) paramagnetic, (III) floating, (IV) antiphase. b) Predicted phase diagram when training the simulated QCNN acting only on the low-bodyness operator subspace. The model is trained on  with $200$ and $20$ states. The crosses mark the training points, while the circle the test ones. Blue circles means correct phase prediction,  while a red color indicates that an incorrect phase was assigned.  }
    \label{fig:ANNNIdiagram}
\end{figure*}

\begin{figure*}[t]
    \centering
    \includegraphics[width=\linewidth]{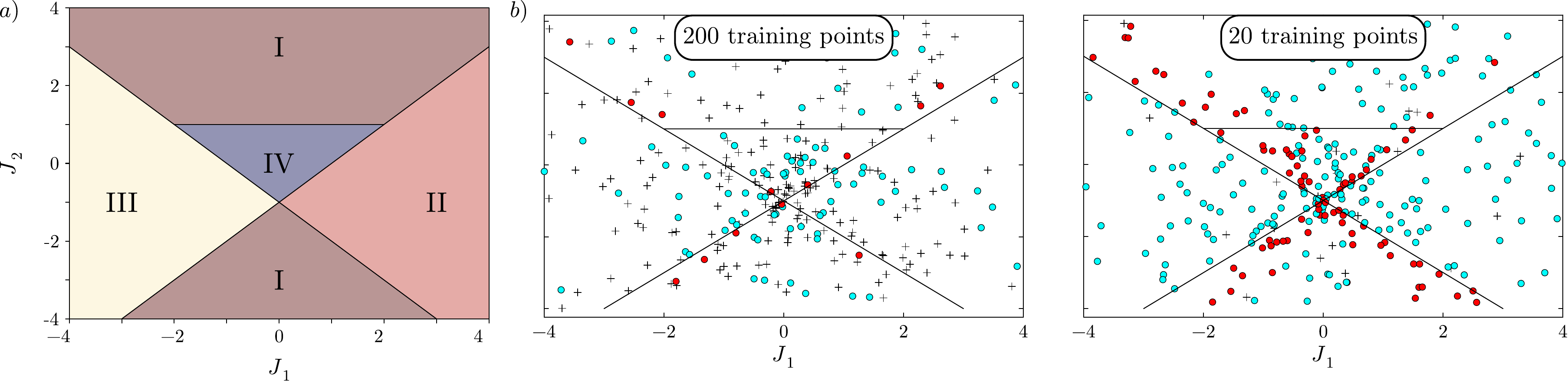}
    \caption{\textbf{Cluster model.} a) Phase diagram for the Hamiltonian in Eq.~\eqref{eq:cluster_model}. The phases are: (I) Haldane, (II) ferromagnetic, (III) anti-ferromagnetic phase, and (IV) trivial. b) Predicted phase diagram when training the simulated QCNN acting only on the low-bodyness operator subspace. The model is trained on  with $200$ and $20$ states. The crosses mark the training points, while the circle the test ones. Blue circles means correct phase prediction,  while a red color indicates that an incorrect phase was assigned.  }
    \label{fig:Clusterdiagram}
\end{figure*}

\subsubsection{Haldane Chain}

Next, we will consider the one-dimensional Haldane chain model~\cite{haldane1983nonlinear} whose Hamiltonian reads
\begin{equation}\label{eq:haldane_model}
H = -J \sum_{i=1}^{n-2} Z_iX_{i+1}Z_{i+2} -h_1\sum_{i=1}^{n}X_i -h_2\sum_{i=1}^{n-1}X_iX_{i+1}  \,.
\end{equation}
Here, we  take $J>0$. As shown in Fig.~\ref{fig:haldane-diagram} the model's phase diagram is characterized by the two ratios $h_1/J$ and $h_2/J$. Inspecting the model in the thermodynamic limit, one can find out the emergence of three distinct phases: antiferromagnetic, paramagnetic and symmetry-protected (SPT). For the classification task, we will focus on  classifying between the paramagnetic phase and the symmetry protected phase, i.e., binary classification, leaving the multi-class classification for the ANNNI model and the cluster Hamiltonian (see below). As such, we fix  $J = 1$ and $h_1 = 0.5$, and set the transition between phases to occur at $h_2 = 0.423$, given by the analysis in the thermodynamic limit~\cite{cong2019quantum}.

Here we consider a chain of $n = 512$ qubits. Throughout the simulation we cap the maximum bodyness of the Heisenberg-evolved measurement operator at three, and we classify the phases only on the expectation values of the $3000$ operators with the largest variance  across the shadows dataset. We use again a dataset of $100$ states and perform training over $200$ iterations, averaged over $5$ different runs. The classification results are shown in Fig.~\ref{fig:XXX-diagram}(b). Similar to the results obtained for the $XXX$ model, we see that, as expected, using larger number of shadows leads to better classification accuracy. In this case, we see that the number of training points has a smaller effect on the model's performance, likely due to the fact that all the states in the same phase have large overlap. Here, at $10$ training points and $500$ shadows per state, our scheme requires only a total of $5000$ measurements, which again constitutes a very small number of shots, as compared to running the QCNN on a quantum computer.
For a more detailed analysis of complexity scaling in the XXX bond-alternating model and the Haldane chain, as well as evidence for the absence of overfitting during training, we refer the reader to Appendix~\ref{ssec:scaling_numerics}.

\subsubsection{ANNNI model}

In this section, we consider the ANNNI model~\cite{elliott1961phenomenological}, which is described by the following Hamiltonian
\begin{equation}\label{eq:ANNNI_model}
H = -J_1\sum_{i=1}^{n-1} X_iX_{i+1} -J_2\sum_{i=1}^{n-2} X_iX_{i+2} -B\sum_{i=1}^{n} Z_i\,.
\end{equation}
The phases of this model are classified in terms of the dimensionless ratios $\kappa = -J_2/J_1$ and $h = B/J_1$. The phase diagram for this model is presented in  Fig.~\ref{fig:ANNNIdiagram}. In particular, we will take $J_1 = 1$ and perform the classification based on the values $\kappa$ and $h$ of Ref.~\cite{cea2024exploring}

In this case we consider a system size of $n= 32$ qubits. Throughout the simulation we cap the maximum bodyness of the Heisenberg-evolved measurement operator at four, where we estimate the expectation of the initial states with non-identity Pauli operators inside a sliding window of size $8$ qubits.  We refer the reader to Appendix~\ref{ssec:methods_PPS} for more details on the truncation. Given that we are now dealing with multiclass classification, we assign phases by measuring two qubits at the QCNN's output and assigning labels using one-hot encoding. Our dataset is now composed of $300$ states and we consider two scenarios where we train over $200$ and $20$ of them (taking  $4000$ and $2500$ shadows per state, respectively). The training is averaged over $5$ different runs. For $200$ training points, we can achieve a train accuracy of $82.8\%$, and a test accuracy of $85.8\%$. Then, for  $20$ training points the train accuracy was of $87\%$, while the test of  $80.2\%$. Thus we can reach similar results to those obtained in the literature, at a much smaller measurement budget.

\begin{table*}[t]
\begin{tabular}{cccccccc}
\cline{2-8}
\multicolumn{1}{l|}{}                     & \multicolumn{1}{l|}{Initial size} & \multicolumn{1}{l|}{Train Points} & \multicolumn{1}{l|}{Test Points} & \multicolumn{1}{l|}{ Amplitude(\%)} & \multicolumn{1}{l|}{TPE(\%)}& \multicolumn{1}{l|}{HEA 1/3 L(\%)} & \multicolumn{1}{l|}{CHE 1/3 L (\%)} \\ \hline

\multicolumn{1}{|c|}{\textbf{MNIST ~\cite{hur2021quantum, oh2020tutorial, baek2022scalable, gong2024quantum, kim2023classical, li2022image, bokhan2022multiclass, ovalle2023quantum, Chang2023Approximately, fan2023hybrid, li2020quantum, zeng2022multi}}
}     &    &    &   &    &  &  & \multicolumn{1}{c|}{}                   \\ \hline

\multicolumn{1}{|c|}{0 / 1}               & \multicolumn{1}{c|}{28x28}        & \multicolumn{1}{c|}{40}           & \multicolumn{1}{c|}{100}         & \multicolumn{1}{c|}{100}                & \multicolumn{1}{c|}{100}   & \multicolumn{1}{c|}{100/98} & \multicolumn{1}{c|}{98/100}              \\ \hline

\multicolumn{1}{|c|}{0 / 9}               & \multicolumn{1}{c|}{28x28}        & \multicolumn{1}{c|}{60}           & \multicolumn{1}{c|}{100}         & \multicolumn{1}{c|}{91}                 & \multicolumn{1}{c|}{99}  & \multicolumn{1}{c|}{93/87} & \multicolumn{1}{c|}{91/91}               \\ \hline

\multicolumn{1}{|c|}{\textbf{Fashion-MNIST ~\cite{hur2021quantum, oh2020tutorial, baek2022scalable, gong2024quantum, kim2023classical, li2022image, bokhan2022multiclass, ovalle2023quantum, Chang2023Approximately}}}  

&    &    &   &    &  & & \multicolumn{1}{c|}{} 

\\ \hline
\multicolumn{1}{|c|}{Trouser / Sweater}   & \multicolumn{1}{c|}{28x28}        & \multicolumn{1}{c|}{60}           & \multicolumn{1}{c|}{100}         & \multicolumn{1}{c|}{98}                 & \multicolumn{1}{c|}{97}   & \multicolumn{1}{c|}{96/95}  &  \multicolumn{1}{c|}{96/93}            \\ \hline
\multicolumn{1}{|c|}{Dress / Sneaker}     & \multicolumn{1}{c|}{28x28}        & \multicolumn{1}{c|}{60}           & \multicolumn{1}{c|}{100}         & \multicolumn{1}{c|}{97}                 & \multicolumn{1}{c|}{100}  & \multicolumn{1}{c|}{99/95}  &  \multicolumn{1}{c|}{98/100}          \\ \hline
\multicolumn{1}{|c|}{\textbf{EuroSAT ~\cite{sebastianelli2021circuit, chalumuri2022quantum, zhang2019polsar, matsumoto2022full, chang2022quantum} }}

 &    &    &   &    &  &  & \multicolumn{1}{c|}{} 
 
 \\ \hline
\multicolumn{1}{|c|}{Industrial / Forest} & \multicolumn{1}{c|}{64x64}        & \multicolumn{1}{c|}{200}          & \multicolumn{1}{c|}{400}         & \multicolumn{1}{c|}{98}               & \multicolumn{1}{c|}{94} & \multicolumn{1}{c|}{94/92}    & \multicolumn{1}{c|}{92$^*$/87$^*$}              \\ \hline
\multicolumn{1}{|c|}{Highway / Sealake}   & \multicolumn{1}{c|}{64x64}        & \multicolumn{1}{c|}{200}          & \multicolumn{1}{c|}{400}         & \multicolumn{1}{c|}{95}               & \multicolumn{1}{c|}{94}  & \multicolumn{1}{c|}{94/92} & \multicolumn{1}{c|}{87/91}              \\ \hline
\multicolumn{1}{|c|}{\textbf{GTSRB ~\cite{aldoski2023impact, li2020quantum, fan2023hybrid, khan2023lightweight}}
}       

&    &    &   &    &  & & \multicolumn{1}{c|}{} 

\\ \hline
\multicolumn{1}{|c|}{20kmh / Stop}        & \multicolumn{1}{c|}{256x256}      & \multicolumn{1}{c|}{120}          & \multicolumn{1}{c|}{120}      & \multicolumn{1}{c|}{98}               & \multicolumn{1}{c|}{97}  & \multicolumn{1}{c|}{96/94}   & \multicolumn{1}{c|}{94/90}             \\ \hline
\multicolumn{1}{|c|}{Bumpy road / Road work}        & \multicolumn{1}{c|}{256x256}      & \multicolumn{1}{c|}{160}          & \multicolumn{1}{c|}{120}         & \multicolumn{1}{c|}{94}               & \multicolumn{1}{c|}{98}   & \multicolumn{1}{c|}{94/90}     & \multicolumn{1}{c|}{86/82}          \\ \hline
\end{tabular}
\caption{\textbf{Results for the classification of classical datasets with a simulated QCNN. }Amplitude refers to amplitude encoding, while angle encodings are denoted by TPE (tensor product encoding), HEA 1/3 (hardware efficient ansatz encoding with one or three layers) and CHE 1/3 (classically hard encoding with one or three layers). All the results are obtained with a constrained-bodyness of 2 operators, except for $^*$ which make use of 3 operators.}
\label{results classical data}
\end{table*}

\subsubsection{Cluster Hamiltonian}

To finish, we consider the cluster model defined by the following Hamiltonian
\begin{equation}\label{eq:cluster_model}
H = \sum_{i=1}^n (Z_i - J_1X_iX_{i+1} - J_2X_{i-1}Z_iX_{i+1})   \,.
\end{equation}
Note that here we employ closed boundary conditions so that $X_{n+1}\equiv X_1 $ and $X_{-1}\equiv X_{n} $. As shown in Fig.~\ref{fig:Clusterdiagram},  the phase transitions are delimited by the ratios between $J_1$ and $J_2$.

Similarly to the previous ANNNI model, we consider a system size of $n= 32$ qubits. Throughout the simulation we cap the maximum bodyness of the Heisenberg-evolved measurement operator at four, where the non-trivial operators are taken from a sliding window of $8$-qubits size. We again assign phases via one-hot encoding (i.e., measure two qubits in the computational basis and assigning a phase depending on what bistring probability is larger). The dataset now consists of $300$ states and we consider training on subsets of $200$ and $20$ of them (taking  $4000$ shadows per state). The training is averaged over $5$ different runs. For $200$ training points, we find a train and test accuracies of $80.9\%$ and $84\%$ respectively. Then, for $20$ training points we obtain a train accuracy of $85\%$, and a test one of $77.6\%$. In all cases, these accuracies are comparable and even higher than the ones achieved in previous attempts for the classification of the Cluster Hamiltonian~\cite{caro2021generalization,gil-fuster2024understanding}.

At this point we find it important to note that for both the ANNNI and cluster models most of the misclassification errors are located near the phase transitions. While this could arise due to the fact that phase transitions are precisely where the classification should become more difficult, we also note that there could be spurious errors arising from the fact that the ``true'' labels in our dataset are assigned by using the thermodynamic limit phase diagram, rather than the  finite size one. This could lead to misalignments between data points and labels that increase as the system size decreases.

\subsection{Classical datasets}

The QCNN has been extensively employed in the literature to showcase the power of QML for classical data classification and characterization. In this section we will show that the most commonly used classical datasets in the literature, are locally-easy. For that purpose, we will simulate the action of the  QCNN for the following datasets: MNIST ~\cite{hur2021quantum, oh2020tutorial, baek2022scalable, gong2024quantum, kim2023classical, li2022image, bokhan2022multiclass, ovalle2023quantum, Chang2023Approximately, fan2023hybrid, li2020quantum, zeng2022multi}, Fashion-MNIST ~\cite{hur2021quantum, oh2020tutorial, baek2022scalable, gong2024quantum, kim2023classical, li2022image, bokhan2022multiclass, ovalle2023quantum, Chang2023Approximately}, EuroSAT ~\cite{sebastianelli2021circuit, chalumuri2022quantum, zhang2019polsar, matsumoto2022full, chang2022quantum} and GTSRB ~\cite{aldoski2023impact, li2020quantum, fan2023hybrid, khan2023lightweight}.

For all the datasets above we will study the simulation of the QCNN with both amplitude and angle encoding. For amplitude encoding, we load each image in the dataset at hand as a square matrix of floating point values, or triplets of them in case of colored data. 
Then, if needed (i.e. for the GTSRB and EuroSAT), we merge the color channels into a single gray-scale matrix. 
After resizing this matrix based on the dataset, it is reshaped as an unnormalized quantum state $\ket{\psi(x)}$ on $n=8$ qubits.
Lastly, we normalize $\ket{\psi(x)}$ and encode it in a Matrix Product State (MPS), by means of sequential singular value decompositions. Then, for angle encoding we will use a tensor product embedding, hardware efficient embedding and the so-called classically hard embedding. We refer the reader to Ref.~\cite{thanasilp2021subtleties} for a detailed description of these data embedding schemes. Importantly, we note that we always take shallow embedding schemes (one and three layers for the hardware efficient and classically hard)  as it is well known that deep versions of these encoding are not useful~\cite{thanasilp2021subtleties}, as they can lead to initial states that are too entangled and thus void the QCNN's trainability guarantees~\cite{pesah2020absence}. A direct consequence of this fact is that all data encoded states will all admit an efficient MPS decomposition. Finally, the simulation of the QCNN is conducted by means of the constrained-bodyness MPS algorithm described in the appendix, and we use a mean-squared error loss function.

In Table~\ref{results classical data} we show results for all the classification tasks considered. Here we can see that the simulated QCNN result in high test accuracies (showing best out 5 independent runs), comparable to, and even larger than, those found in the literature.

Here it is important to note that unlike the quantum dataset case, where the dataset state's preparation could be hard to simulate (thus requiring shadows), all of the QML models used here are fully classically simulable. That is, there is no need to perform measurements on the classical data-encoded states as the embedding scheme itself is classically simulable (this is again a consequence of the results in Ref.~\cite{thanasilp2021subtleties}). As such, it appears that no quantum resources are needed to simulate the QCNN. Concomitantly, this implies that  using QCNN-based QML schemes for classical data appears to be an ill-motivated task.

\section{Discussion}

In this work we have shown, using a proof by demonstration and explicit construction, that one-dimensional tracing out and measurement-based QCNNs can be classically simulated--in the sense that we can construct and efficiently classical surrogates. Our explicit simulations are obtained by conceptualizing the success of QCNNs and showing that they appear to only work on easy problems for which their action can be restricted to polynomially-sized subspaces. Clearly, our work cannot, and does not intend to, prove that there is no scenario whatsoever where it may be necessary to train a QCNN on a quantum computer. However, the burden of proof now rests firmly in the hands of any proponent of QCNNs to identify such cases and until then it is good practice to maintain a healthy skepticism that such cases can be found. Hence we boldly claim:  \textit{There is currently no evidence that QCNNs will work on classically non-trivial tasks, and their place in the upper echelon of promising QML architectures should be seriously revised.} 

With the previous being said, we now present several important caveats to our results. First, while we focus here on the two most popular instantiations of QCNNs used in the literature (one-dimensional tracing out and measurement-based architectures), it is clear that these are the easiest to classically simulate. One could, for instance, envision QCNNs in two or more dimensions, as well as more exotic topologies. Clearly, this would increase the simulation cost, potentially making it prohibitively expensive even at modest sizes. However, even here, Pauli propagation methods, and more general advanced tensor networks techniques (such as projected entangled pair states with belief propagation) can handle these QCNNs~\cite{napp2022efficient}. In this work, we decided not to pursue this route as these QCNNs architectures have not been explored, and are not known to be useful in cases where simple one-dimensional ones fail. As before, we again leave the burden of proof to practitioners to find schemes that avoid classical simulability but also provide real tangible advantages in non-trivial datasets.

Then, it is important to note that our claims do not amount to  a full dequantization of QCNNs. Fundamentally, at least for the case of quantum input data, a quantum computer is needed to obtain classical shadows. However, the resource requirements (state preparation and then single qubit measurements) are substantially easier than running the QCNN on quantum hardware - so much easier that a universal digital quantum computer may no longer be required. 
Thus our results can be be viewed positively as expanding the prospects of the near-term friendly framework of quantum measurement-enhanced machine learning~\cite{huang2021power}.
Another positive spin of our results is a possible setting of ``\textit{classical training and quantum deployment}'', where one trains a QCNN on a classical device via classical shadows, and then uses the optimal parameters to implement and test the model's performance in a quantum computer on new data. This approach has the benefit of bypassing the need to obtain classical shadows from the new data instances. Then, a second positive spin of our work is that we can now start to compare the classical resources needed to perform a simulation, versus the quantum resources needed to actually train a simulable model on a quantum computer. In this ``just because we can classically simulate a model, doesn't mean it is efficient to do so'' perspective, there is ample room for actually deploying  models on quantum computers if the classical cost is  too large. We also leave this open as a future research question.

While we here focused on QCNNs, we remark that the results and lessons learned for this model apply to QML more broadly. We strongly believe that the techniques introduced here can serve as blueprints  to classically simulate the information processing capabilities of other quantum neural networks architectures composed of local parametrized gates. Indeed, one can envision that all of the circuit architectures that are average case simulable via the results in Ref.~\cite{angrisani2024classically}, could also be classically trained via low-bodyness approximations (on locally-easy datasets). Whether there exists good enough solutions within such subspace as they do for QCNNs is an open question which will likely require a case-by-case analysis via direct simulation. 

Finally, we argue that our community is in dire need of non-trivial datasets as our exhaustive literature search did not produce a single example of a task that cannot be classified by simulating the action of the QCNN in the small subspace of low-bodyness observables. Indeed,  for the considered condensed matter quantum datasets, the order parameters are local and allow for classification even through phase transitions. In fact, our results indicate that for classical data, shadow tomography is not even needed, as the whole model is entirely classically simulable. This result pushes back on the hope that the encoding scheme can somehow create classically-hard to simulate features and sheds serious doubt on whether it even makes sense at all to embed classical data on a quantum computer for coherent processing via parametrized quantum circuits such as QCNNs. Whether the fact that we use trivial datasets in our QML model benchmarking is a positive bias effect (i.e., only successful QCNN trainings make it into the published literature) or a more underlying phenomenon of physical problems is left as an open question. Hence, we also put the burden of proof on practitioners to find non locally-easy datasets that quantum models can classify when operating on a quantum computers, but that their classical surrogates cannot solve. Regardless, we believe that an introspection is needed when it comes to choosing QML benchmarking tasks to avoid using trivial ones.    

\section{Acknowledgments}
P. Bermejo, P. Braccia  and L.C. were supported by the Laboratory Directed Research and Development (LDRD) program of Los Alamos National Laboratory (LANL) under project numbers 20230527ECR and 20230049DR. P. Bermejo acknowledges constant support from DIPC. M.S.R. thanks Tyson Jones for helpful suggestions and discussions surrounding the implementation of the LOWESA algorithm. Z.H. acknowledges support from the Sandoz Family Foundation-Monique de Meuron program for Academic Promotion. M.C. acknowledges support from LANL's ASC Beyond Moore’s Law project. 

\bibliography{quantum}

\clearpage
\newpage

\makeatletter
\close@column@grid
\makeatother
\cleardoublepage
\newpage
\onecolumngrid
\renewcommand\appendixname{Appendix}
\appendix

\section*{Appendices}

\section{Proof of Result~\ref{res:bodyness} }

In this section we present a proof for Result~\ref{res:bodyness}. We begin by introducing the basic definitions and theoretical tools needed. 

\subsection{$k$-purities}

Our main goal is to show that the Heisenberg evolved measurement operator essentially only has support in $\OC(1)$-bodyness Paulis. As such, let $\{P_j\}$ denote the set of Pauli operators, and let $k=|P_j|$ denote the bodyness, or weight, of $P_j$, i.e., the number of qubits that $P_j$ acts non trivially on. Given an operator $O\in \BC(\HC)$, the set of bounded operators on the Hilbert space $\HC$, we define its $k$-purity as the projection of $O$ into all the Paulis with bodyness $k$. That is,
\begin{equation}\label{eq:k-puritites}
    p^{(k)}_{O} = \frac{1}{4^n} \sum_{P_j\,:\,|P_j|=k} \Tr [P_j O]^2 \, .
\end{equation}
When $O$ is such that $\norm{O}_2^2=\id$, a condition that we will henceforth assume, then $\sum_{k=1}^n p^{(k)}_{O}=1$ and, since by definition $p^{(k)}_{O}\geq 0$ for all $k$, then the $k$-purities form a  probability distribution. In particular, we are interested in computing the quantities
\begin{equation}\label{eq:average-purity}
    \mathbb{E}_{\thv}\left[p^{(k)}_{\Phi_{\thv}\ad(O)} \right]\,,
\end{equation}
where $\Phi_{\thv}\ad(O)$ denotes the measurement operator obtained from Heisenberg evolving the local measurement through a unitary QCNN as in Eqs.~\eqref{eq:standard_qcnn_conv_pool}. 

Here it is important to note that we have assumed that averaging over $\vec{\theta}$ is equivalent to randomly sampling each local gate in $\Phi_{\thv}$ independently from the Haar measure over $\mathbb{U}(4)$. Then, since the QCNN is unitary according to  Eqs.~\eqref{eq:standard_qcnn_conv_pool}, we can express its action as
\begin{equation}
    \Phi_{\thv}(\cdot)=U(\thv)(\cdot) U\ad(\thv)\,,
\end{equation}
where 
\begin{equation}
    U(\thv)=\prod_l U_l\,.
\end{equation}
Above, each $U_l$ denotes a two-qubit gate in the circuit (see Fig.~\ref{fig:QCNN}(b,left)) and we have omitted the explicit parameter dependence on the right-hand side. Therefore, we have that
\begin{equation}
    \mathbb{E}_{\thv}=\prod_l \int_{\mathbb{U}(4)} d\mu_l(U_l)\,,
\end{equation}
and in order to compute the average $k$-purities we need to integrate local unitaries sampled randomly according to the Haar measure over $\mathbb{U}(4)$. In the next section we will present the basic Weingarten Calculus tools for performing such calculations. 

\subsection{Weingarten calculus}

We here recall a few basic concepts from Weingarten calculus. 
We refer the reader to Ref.~\cite{mele2023introduction,ragone2022representation,garcia2023deep} for additional details.

To begin, let us note that computing the $k$-purities requires evaluating quantities of the form
\begin{align}
  \mathbb{E}_{\thv}\left[p^{(k)}_{\Phi_{\thv}\ad(O)} \right]= \frac{1}{4^n} \sum_{P_j\,:\,|P_j|=k}\mathbb{E}_{\thv}\left[\Tr [P_j U\ad(\thv)OU(\thv)]^2\right]&=\frac{1}{4^n} \sum_{P_j\,:\,|P_j|=k}\prod_l \int_{\mathbb{U}(4)} d\mu_l(U_l) \Tr [P_j^{\otimes 2} (U_l\ad)^{\otimes 2} O^{\otimes 2} U_l^{\otimes 2}]\nonumber\\
   &=\frac{3^k\binom{n}{k}}{4^n} \prod_l \int_{\mathbb{U}(4)} d\mu_l(U_l) \Tr [P_j^{\otimes 2} U_l^{\otimes 2} O^{\otimes 2} (U_l\ad)^{\otimes 2}]\,.
\end{align}
In the last equation we have use the fact that we can always transform one Pauli with a given bodyness onto another one with the same bodyness by local rotations that can be absorbed into the first convolutional layer gate's Haar measure. Above, we take $P_j$ to be any Pauli with bodyness equal to $k$. Importantly, we can vectorize this equation to obtain 
\begin{equation}\label{eq:prod-moments}
  \mathbb{E}_{\thv}\left[p^{(k)}_{\Phi_{\thv}\ad(O)} \right]=\frac{3^k\binom{n}{k}}{4^n}\langle\langle P_j^{\otimes 2}| \prod_l \widehat{\tau}_{l}^{(2)}|O^{\otimes 2}\rangle \rangle\,,
\end{equation}
where we defined the second moment operator for the local $U_l$ gate
\begin{equation}
    \widehat{\tau}_{l}^{(2)}= \int_{\mathbb{U}(4)} d\mu_l(U_l) U_l^{\otimes 2}\otimes (U_l^*)^{\otimes 2}\,.
\end{equation}
Here we recall that the vectorization takes  an operator in $\BC(\HC^{\otimes t})$ and returns a vector in $\HC^{\otimes t}\otimes (\HC^*)^{\otimes t}$ while a channel from $\BC(\HC^{\otimes t })$ to $\BC(\HC^{\otimes t })$ is mapped to a matrix in $\BC(\HC^{\otimes t}\otimes (\HC^*)^{\otimes t})$. Specifically, given some  $X=\sum_{i,j=1}^{d^t} c_{ij}|i\rangle\langle j|$, its vectorized form is $|X\rangle \rangle=\sum_{i,j=1}^{d^t} c_{ij}|i\rangle\otimes | j\rangle$, while given a channel $\Phi(X)=\sum_{\nu=1}^{d^{2t}} K_\nu X J_\nu\ad$, we obtain $\widehat{\Phi}=\sum_{\nu=1}^{d^{2t}}K_\nu \otimes  J_\nu^*$. In particular, the inner product between two vectorized operators is given by $\langle\langle Y|\widehat{\Phi}|X\rangle \rangle=\Tr[Y\ad \Phi(X)]$.

Equation~\eqref{eq:prod-moments} reveals that computing the average $k$-purities requires evaluating the product of the moment operators $\widehat{\tau}_{l}^{(2)}$. These can be computed via the Weingarten calculus as follows. We start by considering a compact unitary Lie group $G$ with Haar measure $d\mu$ acting on a finite-dimensional Hilbert space $\HC$. We are interested in computing the $t$-th fold moment operator, which takes the form: 
\begin{equation}\label{eq:t_twirl}
    \hat{\tau}_{G}^{(t)}(X) = \int_G d\mu(U) U^{\otimes t}\otimes (U^*)^{\otimes t}\,.
\end{equation}
It is well known that the moment operator is a projection onto the (vectorized) $t$-th order commutant of $G$, i.e., the operator vector  space $\textrm{comm}^{(t)}(G)=\{M\in\BC(\HC^{\otimes t })\,|\, [M,U^{\otimes t}]=0\}$, of dimension $d_{G,t}=\dim(\textrm{comm}^{(t)}(G))$. As such, given a basis $\{P_\mu\}_{\mu=1}^{d_{G,t}}$ of $\textrm{comm}^{(t)}(G)$, one can express the  moment operator as 
\begin{equation}\label{eq:twirl-vectorized}
    \widehat{\tau}_{G}^{(t)}=\sum_{\mu,\nu=1}^{d_{G,t}} (W_{G,t}^{-1})_{\nu\mu} | P_\nu\rangle\rangle\langle\langle P_\mu|\,.
\end{equation}
Here $W^{-1}_{G,t}$ is known as the Weingarten Matrix with entries $(W_{G,t})_{\nu\mu}=\Tr[P_\nu \ad P_\mu]$. That is, $W^{-1}_{G,t}$ is the inverse of the commutant's Gram matrix.

We can then proceed to apply this framework to our case of interest. In particular, we will prove the following Lemma (see also~\cite{dalzell2021random,braccia2024computing,napp2022quantifying} for proofs of similar statements). 
\begin{lemma}\label{sup-lemma-1}
   Let $U(\thv)=\prod_l U_l$  be a circuit composed of two qubit gates as in Fig.~\ref{fig:QCNN}(b,left) and assume that each  $U_l$ is sampled independently from a $2$-design over $\mathbb{U}(4)$. Then, the associated second moment operator $\widehat{\tau}_{l}^{(2)}$ associated to each $U_l$ can be represented by the $4\times 4$ dimensional matrix $P$ 
   \begin{equation}
     P = \begin{pmatrix}
        1 & \frac{2}{5} & \frac{2}{5}  & 0\\
        0 & 0 & 0 & 0 \\
        0 & 0 & 0 & 0 \\
        0 & \frac{2}{5}  & \frac{2}{5}  & 1
    \end{pmatrix}\,.
\end{equation}
acting on the subspace spanned by $\{ \ket{i}, \ket{s} \}^{\otimes 2}$, where $\ket{i} \equiv | \id\rangle\rangle $ and $\ket{s} \equiv \vert {\rm SWAP} \rangle\rangle$.   
\end{lemma}

\begin{proof}
let us begin by taking a two-qubit gate $U_l$ acting on qubits $j$ and $j'$. Since $U_l$  is sampled from a $2$-design over $\mathbb{U}(4)$, we recall that the two-fold commutant of the unitary group is $\textrm{comm}^{(2)}(\mathbb{U}(4)) = {\rm span}_{\CC} \{ \id_j \otimes \id_{j'}, {\rm SWAP}_j\otimes {\rm SWAP}_{j'} \}$~\cite{mele2023introduction}. Here  $\id_{j}$ denotes the the identity on the two copies of the $j$-th qubit  Hilbert spaces, while  ${\rm SWAP}_j$ the operation that interchange these two Hilbert spaces (and similarly for $\id_{j'}$ and  ${\rm SWAP}_{j'}$. From Eq.~\eqref{eq:twirl-vectorized} we know that $\widehat{\tau}_{l}^{(2)}:=\widehat{\tau}_{\mathbb{U}(4)}^{(2)}$ will be a projector onto $| \id_j\otimes\id_{j'}\rangle\rangle= | \id_j\rangle\rangle\otimes|\id_{j'}\rangle\rangle$ and $|{\rm SWAP}_j\rangle\rangle\otimes |{\rm SWAP}_{j'}\rangle\rangle$. When two consecutive two-qubit gates act as in Fig.~\ref{fig:QCNN}(b,left), say $U_l$ acting on qubits $j$ and $j'$ and $U_{l+1}$ acting on qubits $j$ and $j''$, we can see that one of the qubits is shared. In this case, $\widehat{\tau}_{l}^{(2)}$ and $\widehat{\tau}_{l'}^{(2)}$ will be projectors onto different subspaces. However, their joint action can be studied by expanding them onto the vector space spanned by the overlaps of their commutants. For instance, $\widehat{\tau}_{l}^{(2)}$ can be fully studied by its action on a four-dimensional vector space spanned by the basis vectors 
\begin{equation}
    \{ \ket{i}, \ket{s} \}^{\otimes 2} \,,
\end{equation}
where $\ket{i} \equiv | \id\rangle\rangle $ and $\ket{s} \equiv \vert {\rm SWAP} \rangle\rangle$. In particular, we can explicitly find that 
\begin{align}
    \widehat{\tau}_{l}^{(2)} \ket{ii} = \ket{ii} \,,\quad 
    \widehat{\tau}_{l}^{(2)} \ket{is} = \frac{2}{5} \ket{is} + \frac{2}{5}  \ket{si} \,,\quad 
    \widehat{\tau}_{l}^{(2)} \ket{si} = \frac{2}{5}  \ket{is} + \frac{2}{5}  \ket{si}\,,\quad
    \widehat{\tau}_{l}^{(2)} \ket{ss} = \ket{ss} \,,
\end{align}
indicating that the second moment, $\widehat{\tau}_{l}^{(2)}$, is given by the  $4 \times 4$ matrix $P$, which we dub $P$-gate,
\begin{equation}\label{eq:matrix-P}
     P = \begin{pmatrix}
        1 & \frac{2}{5} & \frac{2}{5}  & 0\\
        0 & 0 & 0 & 0 \\
        0 & 0 & 0 & 0 \\
        0 & \frac{2}{5}  & \frac{2}{5}  & 1
    \end{pmatrix}\,.
\end{equation}
\end{proof}

Combining Eqs.~\eqref{eq:prod-moments} and~\eqref{eq:matrix-P} we find that the average $k$-purity can be obtained by projecting $|O^{\otimes2}\rangle\rangle$ and $|P_j^{\otimes 2}\rangle\rangle$ into the $i$ and $s$ basis and evolving them through a circuit composed of $P$ gates respecting the same topology as that of the QCNN. We refer the reader to Ref.~\cite{braccia2024computing,dalzell2022randomquantum,napp2022quantifying} for additional details on this general procedure.

\subsection{Exact derivation of the $k$-purities for a prototypical QCNN ansatz}\label{prototypical QCNN}

\begin{figure}[h]
    \centering
    \includegraphics[width=0.6\linewidth]{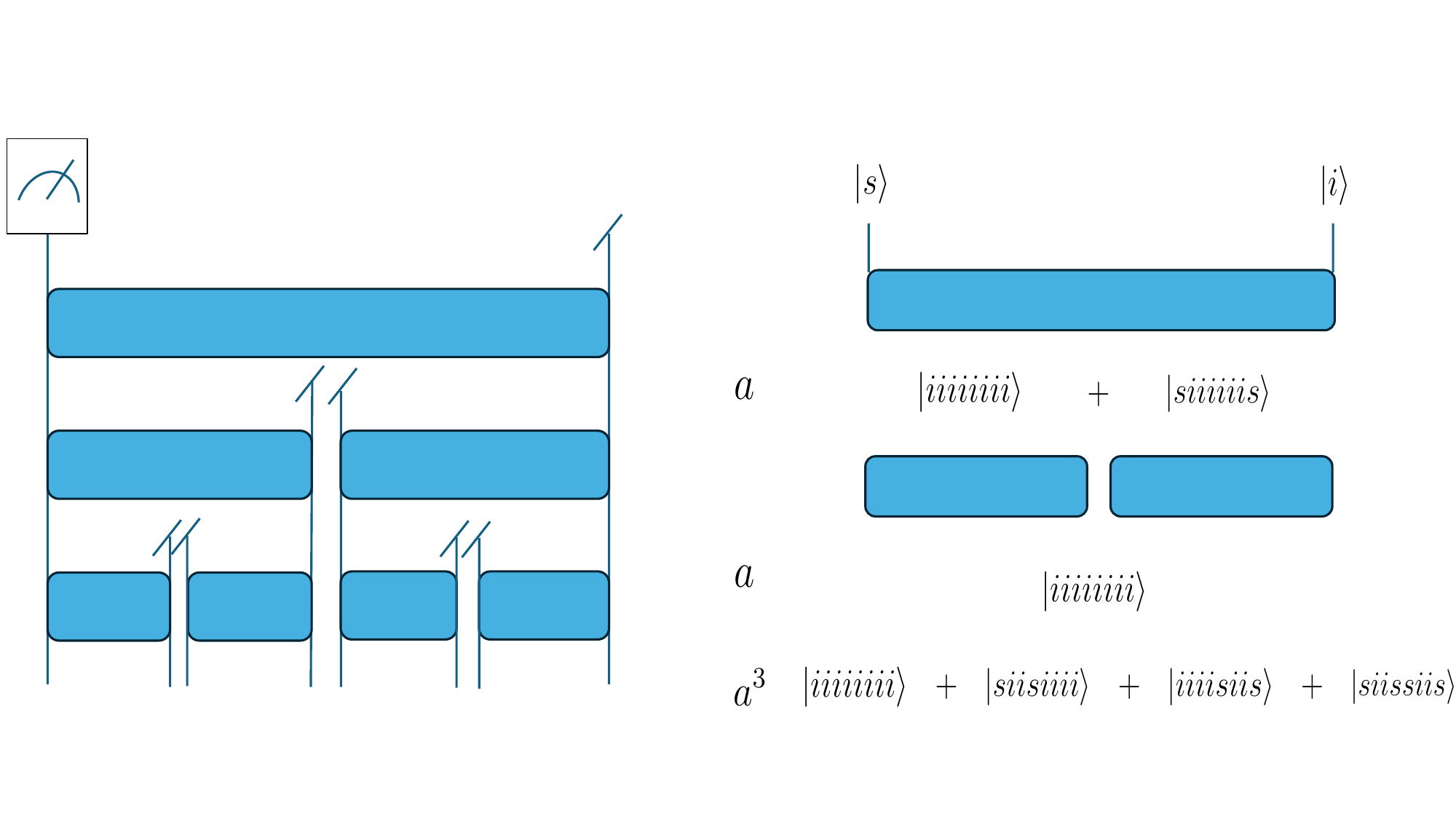}
    \caption{\textbf{Schematic representation of the proof technique.} In this figure we show a QCNN architecture (left) and the spread of the vectorized measurement operator as it propagates through the $P$ gates of the first 2 layers of the QCNN. The initial state is given by $\ket{s}$ on the first qubit and $\ket{i}$ on the remaining qubits.  Here, we also include the coefficient $a$ defining the contribution of each operator to the $k$-purities, which is obtained after applying the $P$-gate on each gate and equals $2/5$, according to Eq.~\eqref{eq:matrix-P}. Note that the QCNN on the left is exactly the same as that in Fig.~\ref{fig:QCNN} (b,left) up to an unimportant swapping of the qubits at the output of each unitary in the convolutional layers.}
    \label{qcnn_bodyness}
\end{figure}

To begin, let us recall the claim of Result~\ref{res:bodyness}:
\begin{result}[Informal]\label{res:bodyness-SI} 
    Consider QCNNs where the convolutional layers are composed of general parametrized two-qubit gates acting on nearest neighboring qubits in a pattern such as those of  Fig.~\ref{fig:QCNN} (b,left). In average, the contribution in $\Phi_{\vec{\theta}}\ad(O)$ of a given Pauli with bodyness $k$, decays exponentially with $k$.
\end{result}

In what follows, we will prove the following theorem, which constitutes a formal version of the previous claim:

\begin{theorem}\label{theo-sup}
    Consider a QCNN acting on $n=2^\eta$ qubits, with $\eta\in\mathbb{N}$, as in Fig.~\ref{fig:QCNN} (b,left), or alternatively as in Fig.~\ref{qcnn_bodyness}(b,left). That is, 
\begin{equation}
    \Phi_{\vec{\theta},\vec{\lambda}} =  \bigcirc_{l=1}^{\eta-1} \left(P_{l}^{\vec{\lambda}_{l}} \circ C_{l}^{\vec{\theta}_{l}}\right),
\end{equation}
where $P_{l}^{\vec{\lambda}_{l}}(\cdot)=\Tr_{S_l}[\cdot]$ is a pooling layer where the qubits in the subset $S_l$ are traced out and where $C_{l}^{\vec{\theta}_{l}}=U_l(\theta_l)(\cdot)U_l(\theta_l)$ is a convolutional layer such that
\begin{equation}
   U_l(\vec{\theta}_l)=\bigotimes_{t=1}^{n_l/2} U_l^t (\vec{\theta}_l^t),
\end{equation}
where $U_l^t$ is a two-qubit gate acting on qubits $2t$ and $2t-1$. Above, $n_l=1,\dots,2^{\eta-l}$ and $S_l=\{1,3,5,\cdots,\eta-1\}$. Then, assuming that $U_l^t$ forms an independent $2$-design over the unitary group $\mathbb{U}(4)$ and denoting the average $\mathbb{E}_{\vec{\theta}}=\prod_{l=1}^{\eta-1}\prod_{t=1}^{n_l/2}\mathbb{E}_{\vec{\theta}_l^t}$ one obtains that

\begin{equation}\label{eq:k-pur-2}
\begin{aligned} 
    \mathbb{E}_{\thv}\left[p^{(k)}_{\Phi_{\thv}\ad(O)} \right] = \frac{2}{3} \left(\frac{3}{2}\right)^k\sum_{k_{L}^s=1}^{2k_{L-1}^s}\left(\frac{1}{2}\right)^{2k_L^s-k}
    \sum_{k_{L-1}^s=1}^{2k_{L-2}^s}...\sum_{k_3^s=1}^{2k_2^s}\sum_{k_2^s=1}^{2k_1^s}\binom{2k_1^s}{k_2^s}\binom{2k_2^s}{k_3^s}...\binom{2k_{L-1}^s}{k_L^s} a^{1+2\sum_{l=1}^{L-1}k_l^s}\,,
\end{aligned}
\end{equation}
where $O$ is a single qubit Pauli operator, and $p^{(k)}_{\Phi_{\thv}\ad(O)}$ is the $k$-purity as defined in Eq.~\eqref{eq:k-puritites}.
\end{theorem}

We proceed to apply the machinery introduced in the previous section to compute the average $k$-purities of an operator that is Heisenberg evolved throughout the QCNN architecture  shown in Fig.~\ref{qcnn_bodyness}(b,left). 
By construction, the gates in each layer of the QCNN do not cross, i.e. they act on independent pairs of qubits. We choose to study this particular configuration because it reduces the mixing of operators, keeping the exact calculation of the operator bodyness tractable. Moreover, it provides a reliable low-bound for the operator mixing in alternative QCNN architectures.  When the convolutional layer is composed of two layer of single-qubit gates as in Fig.~\ref{fig:QCNN}(b,right), then the proof of  Result~\ref{res:bodyness} can be found in~\cite{pesah2020absence}. 

\begin{proof}

As previously mentioned, the measurement operator $O$ is assumed to be a single Pauli operator (acting e.g., on the topmost remaining qubit). As per Lemma~\ref{sup-lemma-1}, we are required to work in the $\{\ket{i}, \ket{s}\}^{\otimes n}$ basis, so we will rewrite this operator as $|O^{\otimes 2}\rangle\rangle = \frac{2}{3} \ket{s}\ket{i}^{\otimes n-1}$. 
Given that each gate on our ansatz maps $\ket{ii}$ to $\ket{ii}$, we are interested in studying the propagation of $\ket{si}$ throughout the QCNN, spreading from $\ket{s}\ket{i}^{\otimes n-1}$. In Fig.~\ref{qcnn_bodyness}, we exemplify the spread of $\ket{si}$ throughout the first 2 layers of the QCNN, keeping track of the coefficients picked up from each gate.

By construction, the topology of our QCNN ansatz enforces each $P$-gate to act on, at least, one $\ket{i}$. Hence, we can restrict our study to the remaining input operator, which can be either $\ket{s}$ or $\ket{i}$.
We are interested in characterizing the support of the output of the network of $P$-gates over the non trivial components of $\{\ket{i},\ket{s}\}^{\otimes n}$, grouped by their $\ket{s}$ content. Noticing that each operator $\ket{s}$ entering a $P$-gate will either generate a trivial output $\ket{ii}$ or a non-trivial one $\ket{ss}$, we deduce that, after each layer, the number of $\ket{s}$ terms appearing in each possible basis state reached by our ansatz is even. Thus, we now introduce a variable $k_l^s$, which we refer to as the $s$-content, counting the number of pairs of $\ket{s}$ operators appearing in the output states after layer $l$ is applied.
It is easy to see that given a state with $k_{l-1}^s$ its outputs after layer $l$ will contain all $k_l^s$ from zero, corresponding to each gate mapping its $\ket{si}$ input to $\ket{ii}$, to $2k_{j-1}^s$, corresponding to the case where all the outputs are $\ket{ss}$. Each of the possible states on layer $l$ with $s$-content $k_l^s$ appears exactly $\binom{2k_{l-1}^s}{k_l^s}$ times.

Assuming the number of qubits $n$ is such that the QCNN comprises $L$ layers, we can then calculate all the possible appearances of non-trivial basis states with fixed $s$-content $k_L^s$ at the end of the circuit, $N(k_L^s)$, as:
\begin{equation}
\begin{aligned}
    N_p(k_L^s) &= \sum_{k_{L-1}^s=1}^{2k_{L-2}^s}...\sum_{k_3^s=1}^{2k_2^s}\sum_{k_2^s=1}^{2k_1^s}\binom{2k_1^s}{k_2^s}\binom{2k_2^s}{k_3^s}...\binom{2k_{L-1}^s}{k_L^s}=\sum_{k_{L-1}^s=1}^{2k_{L-2}^s}...\sum_{k_3^s=1}^{2k_2^s}\sum_{k_2^s=1}^{2k_1^s} n(k_1^s,\dots,k_{L}^s)\,,
\end{aligned}
\end{equation}
subjected to $k_1^s=1$. This constraint follows from the initial state $\ket{s}\ket{i}^{\otimes n-1}$ only branching into a state with a single $\ket{ss}$ and the completely trivial state (which we discard for the purpose of the analysis). Notice that we also introduced the number of configurations with a fixed evolution of the $s$-content $n(k_1^s,\dots,k_{L}^s)$.
Now that we have classified the support of the average Heisenberg evolved $|O^{\otimes 2}\rangle\rangle$ in the basis $\{\ket{i},\ket{s}\}^{\otimes n}$ by its $s$-content $k_L^s$ at the end of the circuit, we can proceed to analyze the contribution of each of these terms to the distribution of $k$-purities. First of all, we need to incorporate the coefficients they pick up when each gate is applied. Considering that the first layer always generates a coefficient $a=2/5$, and each $k_{l>1}^s$ does also carry a multiplicative factor $a^{2k_l^s}$, each operator at the end of the circuit will pick up a coefficient  $a^{m(k_1^s,\dots,k_{L-1}^s)}$ depending on the path it followed thorough the network of $P$-gates, where the exponent $m(k_1^s,\dots,k_{L-1}^s)$ is obtained as:
\begin{equation}
m(k_1^s,\dots,k_{L-1}^s) = 1+2\sum_{l=1}^{L-1}k_l^s \,.
\end{equation}
Lastly, we map the $s$-content to Pauli bodyness and obtain the $k$-purities. To do this, we recall that the operator $\ket{s}$ is defined as $\ket{s}$ = $\frac{\vert \id^{\otimes 2}\rangle\rangle + \vert X^{\otimes 2}\rangle\rangle + \vert Y^{\otimes 2}\rangle\rangle + \vert Z^{\otimes 2}\rangle\rangle}{2}$, which we can schematically rewrite as $\ket{s}$ = $\frac{1}{2}\vert \id^{\otimes 2}\rangle\rangle + \frac{3}{2}\vert P^{\otimes 2}\rangle\rangle$ since there is no distinction between Paulis $P$ in the analysis of Pauli weight.
We can thus see that each state with $s$-content $k_L^s$ splits into $k$-bodied Paulis where $k$ goes from zero to $2k_L^s$, and where one associates to each of these Paulis a coefficient $(\frac{3}{2})^k$ $(\frac{1}{2})^{2k_L^s-k}$, since the remaining identities $\ket{i}$ do also map to identities $|\id^{\otimes 2}\rangle\rangle$ and pick up no coefficient.

Putting everything together, and reintroducing the initial coefficient $\alpha=2/3$, we finally arrive at the exact expression for our average QCNN $k$-purities
\begin{equation}\label{eq:k-pur}
\begin{aligned} 
    \mathbb{E}_{\thv}\left[p^{(k)}_{\Phi_{\thv}\ad(O)} \right] &= \frac{2}{3} \left(\frac{3}{2}\right)^k\sum_{k_{L}^s=1}^{2k_{L-1}^s}\left(\frac{1}{2}\right)^{2k_L^s-k}
    \sum_{k_{L-1}^s=1}^{2k_{L-2}^s}...\sum_{k_3^s=1}^{2k_2^s}\sum_{k_2^s=1}^{2k_1^s}n(k_1^s,\dots,k_L^s) a^{m(k_1^s,\dots,k_{L-1}^s)}\\
    &= \frac{2}{3} \left(\frac{3}{2}\right)^k\sum_{k_{L}^s=1}^{2k_{L-1}^s}\left(\frac{1}{2}\right)^{2k_L^s-k}
    \sum_{k_{L-1}^s=1}^{2k_{L-2}^s}...\sum_{k_3^s=1}^{2k_2^s}\sum_{k_2^s=1}^{2k_1^s}\binom{2k_1^s}{k_2^s}\binom{2k_2^s}{k_3^s}...\binom{2k_{L-1}^s}{k_L^s} a^{1+2\sum_{l=1}^{L-1}k_l^s}\, .
\end{aligned}
\end{equation}

\end{proof}

\begin{figure}[h]
    \centering
    \includegraphics[width=0.6\linewidth]{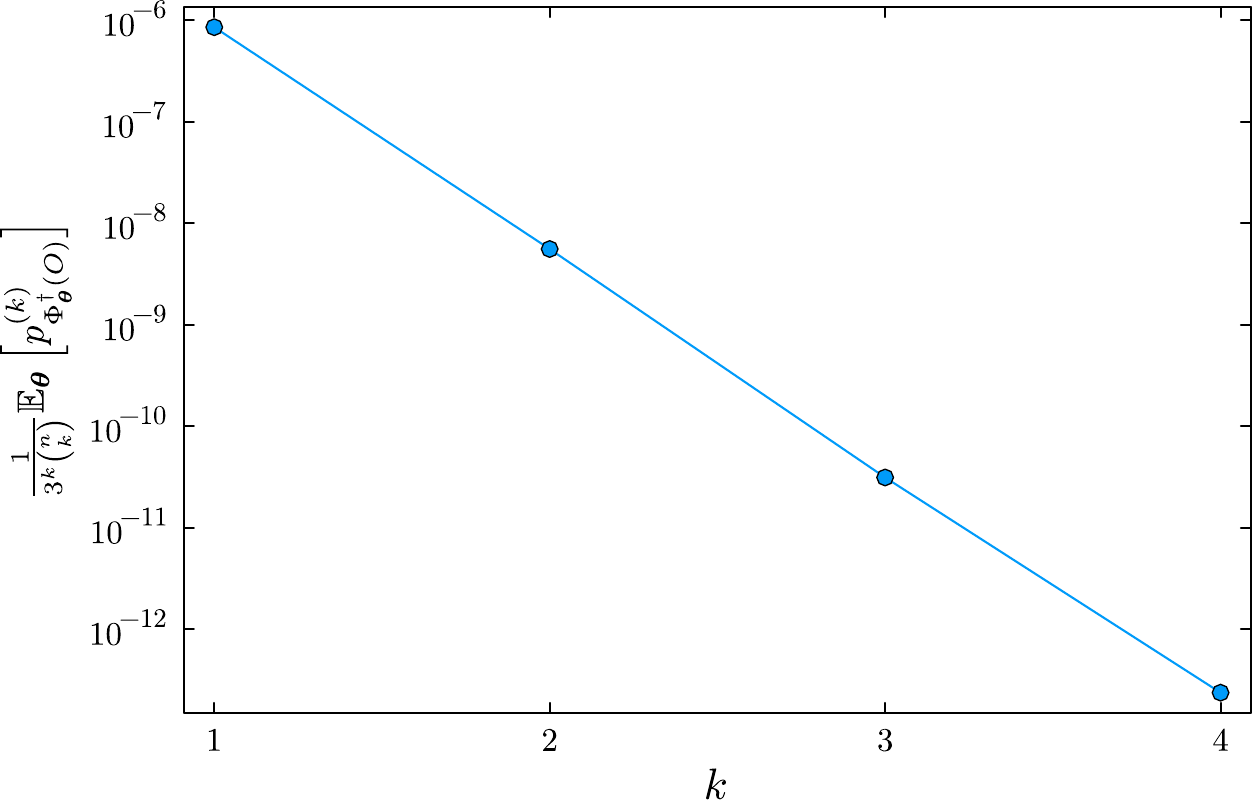}
    \caption{\textbf{Average contribution per bodyness for different problem sizes.} In this figure we plot $\frac{1}{3^k\binom{n}{k}}\mathbb{E}_{\thv}\left[p^{(k)}_{\Phi_{\thv}\ad(O)} \right]$ for a system of $n=128$ qubits. This quantity determines the average contribution of a given Pauli with bodyness $k$ on the Heisenberg evolved measurement operator.  }
    \label{fig:purities}
\end{figure}

We can  evaluate Eq.~\eqref{eq:k-pur} and plot the ensuing results in Fig.~\ref{fig:purities} where we plot the average contribution of each Pauli with bodyness $k$ to the $k$-purities (i.e., we plot  $\frac{1}{3^k\binom{n}{k}}\mathbb{E}_{\thv}\left[p^{(k)}_{\Phi_{\thv}\ad(O)} \right]$). Here we can see that the  Heisenberg evolved measurement operator through a randomly initialized QCNN has support essentially only on Paulis with bodyness in $\OC(1)$. Indeed, we can also verify via the results in Ref.~\cite{braccia2024computing} that a QCNN with a topology such as that in Fig.~\ref{fig:QCNN}(b,right) follows exactly the same behavior.

\section{Classical simulability of QCNNs}\label{apx:classical_methods}

Having realized that the success of QCNNs arises from the fact that they are initialized, explore and  end their training in the polynomially-large subspace of low-bodyness observables, we proceed to describe methods to classically simulate their action on this subspace. All techniques are based on the idea of only processing the information encoded in low-bodyness measurements of the input states. In particular, the methods presented here truncate the Heisenberg-evolved measurement operators to low bodyness (i.e., bodyness in $\OC(1)$), which allows us to guarantee an efficient representation of this backwards-evolved operator. Then, the overlap with the initial states are obtained by projecting each $\rho_i$ into the low-bodyness subspace (e.g., by first taking Pauli classical shadows in the case of quantum datasets). The first simulation method we use is based on the LOWESA algorithm~\cite{fontana2023classical,rudolph2023classical}, which belongs to the class of \textit{Pauli Propagation Surrogates}, within the family of \textit{Pauli Propagation} algorithms. For the second method, we employ tensor networks with restricted bodyness.

\subsection{LOWESA 
}\label{sec:methods_lowesa}
To introduce the novel variant of LOWESA, which we classify as a \textit{Pauli Propagation Surrogate} (PPS), we begin by giving an overview of the underlying \textit{Pauli Propagation} (PP) simulation. Then we will move on to the PPS we use for QCNN classification.

\subsubsection{\textbf{Pauli propagation}}\label{ssec:methods_PP}
\textit{Pauli Propagation} (PP) refers to the simulation of an expectation value in the form
\begin{equation}\label{eq:general_expectation}
    \langle O \rangle = \Tr[U(\thv)\rho U^\dagger(\thv)O] = \Tr[\rho U^\dagger(\thv)OU(\thv)]\,,
\end{equation}
via a Pauli path integral approach~\cite{aharonov2022polynomial}. Specifically, we apply the parametrized quantum circuit $U(\thv)$ to the observable $O$ in the Pauli Transfer Matrix (PTM) formalism~\cite{chow2012universal} and then individually compute the trace of each resulting Pauli operator with the initial state $\rho$.  We highlight that our following PP notation assumes \textit{normalized Pauli operators} such that $\Tr[P_i^2] = 1$. 

First, let us define 
\begin{equation}
    U[\cdot] \coloneqq U^\dagger(\cdot)U
\end{equation}
as the action of an operator in the PTM formalism. Furthermore, we define the quantum circuit as 
\begin{equation}
    U(\thv) = \prod_{i=1}^{m} U_i(\theta_i)\,,
\end{equation}
where $U_i(\theta) =  e^{-i \theta P_i}$ is a unitary generated by the Pauli operator $P_i$ with parameter $\theta_i$. The circuit could also contain Clifford operators like $H$ or CNOT, or channels like Pauli noise that are diagonal in PTM representation at very low computational cost. However, the examples of QCNNs we consider in this work can be fully rewritten in terms of Pauli gates. 

We can write the action of a Pauli gate on a (normalized) Pauli operator $P_j$ as 
\begin{align}
    U_i(\theta)[P_j] &= e^{i \theta P_i}P_j e^{-i \theta P_i}\\
         &= \left(\cos(\frac{\theta}{2})I + i \sin(\frac{\theta}{2})P_i\right) P_j \left(\cos(\frac{\theta}{2})I - i \sin(\frac{\theta}{2})P_i\right) \,.
\end{align}
This general formula simplifies depending on whether $P_i$ and $P_j$ commute (i.e., $[P_i, P_j] = 0$) or anti-commute (i.e., $\{P_i, P_j\} = 0$). Using trigonometric identities, we find that
\begin{equation}
    U_i(\theta)[P_j] = \begin{cases}
        P_j, &\text{ if } [P_i,P_j] = 0\\
        \cos(\theta)P_j - i\sin(\theta)P_iP_j, &\text{ if } \{P_i,P_j\} = 0\,.
    \end{cases}
\end{equation}
With the product of Pauli operators given as $P_iP_j = i \epsilon_{ijk}P_k$ for $P \neq I$, we see that the application of a Pauli gate to a Pauli operator either leaves the Pauli operator unchanged, or it creates two different Pauli operators with \textit{real-valued} trigonometric coefficients. This implies that simulating such a quantum circuit can create a number of Pauli operators that is exponential in the number of gates $m$. 

After applying the entire quantum circuit, we receive a weighted sum of Pauli operators,
\begin{align}
    U(\thv)[O] = \sum_{\alpha} c_\alpha(\thv) P_\alpha \,,
\end{align}
where $c_\alpha(\thv)$ are trigonometric polynomials. This expression goes over \textit{unique} Pauli operators because we \textit{merge} Pauli paths, i.e., add their coefficients if Pauli operators become equal at some stage of the quantum circuit. 
Using this Pauli path formulation, the original expectation function in Eq.~\eqref{eq:general_expectation} becomes
\begin{align}
    \langle O \rangle &= \Tr[\rho U^\dagger(\thv)OU(\thv)]\\
    &= \sum_\alpha c_\alpha(\thv) \Tr[\rho P_\alpha]\\ 
    &= \sum_{\beta}\sum_\alpha c_\beta c_\alpha(\thv) \Tr[P_\beta P_\alpha]
    \,,\label{eq:PP_expectation}
\end{align}
where $\rho = \sum_{\beta}c_\beta P_\beta$ is the decomposition of the initial state in the Pauli basis.

We can restrict the simulation to the polynomially small operator subspace of low-body operators by truncating a Pauli path if the propagating Pauli operator crosses a truncation threshold. This strongly reduces the sum over $\alpha$ to a classically manageable subset of low-body Pauli operators.

The crucial insight in Eq.~\eqref{eq:PP_expectation}, which leads to the PP surrogates, is that the initial states $\rho$ are known at the time of simulation. This means that both the magnitude of the coefficients $c_\beta$ and overlaps $\Tr[P_\beta P_\alpha]$ could in principle be computed before starting the optimization of the parameters $\bm\theta$. One could therefore invest pre-computation time to identify which indices $\beta$ are most important to the task at hand, and only evaluate along the Pauli paths leading to those Pauli operators $P_\beta$. This is the idea behind PP surrogates.

\begin{figure}
    \centering
    \includegraphics[width=0.6\linewidth]{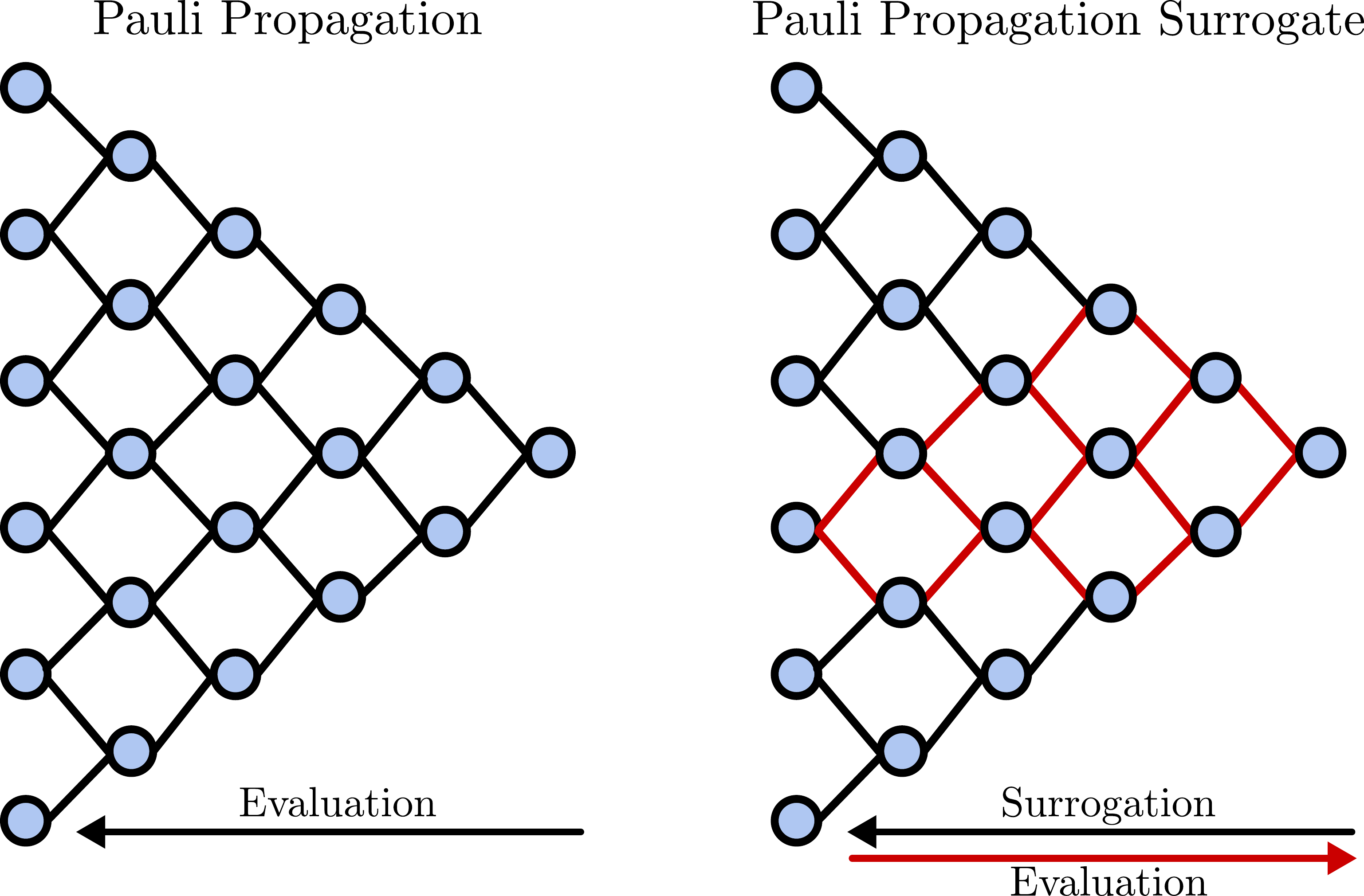}
    \caption{\textbf{Schematic comparison between Pauli Propagation (PP) and Pauli Propagation Surrogate (PPS) methods.} Both methods are based on propagating Pauli operators through a quantum circuit which will in general create new Pauli operators via splitting of paths. If we employ a \textit{breadth-first} approach to the propagation, we can identify identical Pauli operators and merge them. The splitting and merging pattern depicted here is strongly simplified. In conventional PP, one would numerically evaluate the coefficients throughout the circuit and then calculate the overlap with the initial state to estimate the expectation value. In contrast, PPS first compute a graph representation of the Pauli paths, which we call the \textit{surrogation} step. Evaluation of the expectation value is then performed only along the paths that result in operators that are deemed important for the task.}
    \label{fig:PP_and_PPS}
\end{figure}

\subsubsection{\textbf{Pauli propagation surrogate}}\label{ssec:methods_PPS}

\textit{Pauli Propagation Surrogates} (PPS) are a family of PP algorithms that trade pre-compute time and memory for drastically faster evaluation of the expectation function. As such, PPS algorithms can be classified as \textit{classical surrogates}~\cite{schreiber2022classical}. The PPS algorithm we employ in this work is a novel variant of the \textit{low-weight efficient simulation algorithm} (LOWESA) presented in Refs.~\cite{fontana2023classical,rudolph2023classical}. As hinted at in the previous section, instead of numerically evaluating all coefficients $c_\alpha$ during the \textit{Pauli Propagation}, we only store a graphical representation of the \textit{splitting} and \textit{merging} of Pauli paths during an initial \textit{surrogation} pass. We then identify which paths lead to the most important $P_\beta$ operators and only re-evaluate the computed graph along those paths. A schematic depiction of this is shown in Fig.~\ref{fig:PP_and_PPS}.

Consider the simple but common case of 
    $\rho=|0\rangle\langle 0| = \left(I+Z\right)^{\otimes n}$ ,
where the individual traces $\Tr[\rho P_\alpha] \in \{0, 1\}$ are either 0 or 1 depending on if $P_\alpha$ contains $X, Y$ operators on any qubit or not, respectively. Again, note the fact that we are using normalized Pauli operators and can thus drop the $\frac{1}{2}$ factor per qubit. The number of operators that contain only $I, Z$ operators is generally much lower than the number of all propagated Pauli operators, but only they contribute to the expectation value. This means that the graph representation of the expectation value can be re-evaluated only on the paths leading to those operators, resulting in a significant speed-up.

In this work, we do not consider such simple initial states but instead classical shadow representations~\cite{huang2020predicting} of more intricate quantum states. That is, we use measurements of $\rho$ in random Pauli bases to estimate the values $c_\beta = \Tr[\rho P_{\beta}]$. Replacing $\rho$ with its classical shadow $\rho_s = \sum_\beta c_{\beta,s} P_\beta$, we receive the expectation estimator
\begin{align}\label{eq:PP_expectation_approximation}
    \langle O \rangle &\approx\sum_{\alpha} c_{\alpha}(\thv) \Tr[\rho_s P_{\alpha}] \\
    &= \sum_{\beta}\sum_\alpha c_{\beta,s} c_\alpha(\thv) \Tr[P_\beta P_\alpha]
    \,.\label{eq:PP_shadow_expectation}
\end{align}

\emph{Propagation Truncations} -- To maintain the efficieny of the simulation, we employ truncations of the Pauli propagation during the initial surrogation. First, we restrict the propagation to low-bodyness operators. Depending on the difficulty of the problem, we choose a truncation threshold $k$, and if a Pauli operator crosses this threshold at any part of the circuit, it is not propagated further. As such, the operators $P_\alpha$ in Eq.~\eqref{eq:PP_expectation_approximation} and~\eqref{eq:PP_shadow_expectation} have at most this bodyness $k$. This truncation is proven to be highly effective for average-case simulability in the accompanying Ref.~\cite{angrisani2024classically}. Furthermore, we employ a so-called \textit{frequency} truncation which affects the coefficients $c_\alpha(\thv)$. We define the frequency of a path as the number of sines and cosines it has picked up. The average coefficient of a path with $\ell$ sines and cosines over the entire parameter landscape is $\left(\frac{1}{2}\right)^\ell$, which gives frequency truncation the intuitive meaning of an average coefficient truncation. Both truncations have been used in Ref.~\cite{rudolph2023classical} with good practical success.

\emph{Surrogate Truncations} -- While the complexity of the initial surrogation can be controlled by the propagation truncation above, PP surrogates allow further truncations that speed up expectation and gradient evaluations. First, we do not always choose to estimate the expectation values of all operators $P_\beta$ up to bodyness $k$. This would imply that we expect the underlying correlations in the training data to be \textit{all-to-all}. Instead, we can choose to evaluate operators that only contain non-identity Pauli operators in a \textit{sliding window} between 1D neighboring qubits, i.e, non-identity Pauli operators are taken from a subset of adjacent qubits, which is displaced along the qubit chain. If the data is 1D structured, non-trivial Pauli operators on adjacent qubits are expected to carry most of the discriminating information. Other subsets of low-bodyness operators can be chosen on a case-by-case basis. Another strategy we employ to sparsify the evaluation graph is based on the variance over the input states. We can estimate the variance of the coefficients $c_{\beta, s}$ across the dataset (potentially inside the sliding window) and only evaluate along the paths with high variance. This leverages the heuristic that high-variance paths are likely the ones that are most useful for classifying the states, as opposed to paths that have the same coefficient irrespective of the class label.

\subsection{Constrained bodyness tensor network}

\label{constrained TN}
Tensor networks stand out for their convenient properties to represent vector states based on their local features. Moreover, they serve as a suitable platform to efficiently perform certain computations that would otherwise result in prohibitive computational costs. This is exemplified in a recent study of random quantum circuits \cite{braccia2024computing}, where the authors introduced an efficient Matrix Product State (MPS) based algorithm to perform the projection of a vectorized operator into the subspace of $k$-bodyness Paulis. This setting paves the way to control the bodyness of the operators in an efficient manner, enabling tensor networks to deal with polynomially-sized operator subspaces and simulate efficiently a wide variety of ans\"{a}tze under this operator constraint. In this work, we made use of this construction to show that the classical datasets considered in the main text are indeed classifiable with only access to the reduced subspace of low-bodyness operators during the entire training.

This method is currently applicable to binary classification tasks and fares best with one-dimensional quantum systems, we leave for future work the extension to multi-class problems and higher dimensional topologies.
We now proceed to explain the details of the technique.

In outline, the method consists of three steps. First we pre-process the input data, be it quantum or classical, in such a way that only the low-bodyness information is kept. Then, we solve the binary classification task by finding the optimal classifying operator as represented by an MPS in the low-bodyness operator subspace. Lastly, we compile the quantum circuit ansatz at hand such that the Heisenberg evolved measurement operator is as close as possible to the one found at the previous step.

The pre-processing of the input data works as follows. Consider we are given a set of quantum states $\{\rho_i\}$, which could be either the result of some purely quantum experiment or the result of encoding classical data into a quantum computer. We can resort to shadow tomography techniques ~\cite{huang2020predicting,sauvage2024classical, huang2021provably} to efficiently reconstruct the components of each state over the subspace of small-weight Paulis, resulting in a truncated, classical dataset $\{\tilde{\rho}_i\}$. Of course, when the data is classical to start with, we can straightforwardly encode it into the low-body space. For instance, one can simulate the encoding circuit via MPS methods and later project the resulting states via the projectors $\phi_k$ introduced in \cite{braccia2024computing}. 
No matter the path taken, we then vectorize the states $\tilde{\rho}_i$ and normalize them to obtain a new dataset $| \tilde{\rho}_i \rangle \rangle$ where the truncated density matrices are encoded as quantum states in MPS form. Notice that in doing this, the physical dimension of the quantum systems constituting the quantum computational register goes from $d$ to $d^2$, namely, for the qubits systems we deal with, from 2 to 4.

In this formalism quantities such as $\Tr [\rho O]$ can be readily expressed as inner products $\langle \langle \rho | O \rangle \rangle$, which in turn are easy to compute when the states involved can be efficiently represented as MPSs. Hence, we can carry out the training by optimizing overlaps between MPSs, where the particular optimization scheme depends on the loss function chosen to guide the learning task. As an example, we can formalize this description in the case of the mean squared error, which is a common loss function for supervised learning tasks, and the one used in the problems considered in this manuscript. We recall that the mean squared error loss function is defined by
\begin{equation*}
MSE(\theta) = \frac{1}{N_t}\underset{i=1}{\overset{N_t}{\sum}}(y_i - \Tr(\tilde{\rho_i}U^{\dag}(\vec{\theta})MU(\vec{\theta})))^2 \,.
\end{equation*}
Here, the sum runs over the whole training set, $\tilde{\rho_i}$ stands for the  density matrix representation of our truncated $i$-th training sample, $y_i$ stands for the known $i$-th label, $M$ is the measurement operator and $U^{\dag}(\vec{\theta})(\cdot)U(\vec{\theta})$ represents the Heisenberg evolved operator parametrized by the parameters $\vec{\theta}$. Using our MPS setting, one would reformulate this expression as:
\begin{equation*}
MSE(|O\rangle\rangle) = \frac{1}{N_t}\underset{i=1}{\overset{N_t}{\sum}}(y_i - \langle\langle \tilde{\rho_i} \vert O \rangle\rangle)^2 \,,
\end{equation*}
where $ \vert O \rangle \rangle$ replaces the Heisenberg evolved operator $\vert U^{\dag}(\vec{\theta})MU(\vec{\theta}) \rangle \rangle$ in the vectorized picture, and where now the optimization is carried out on the tensors defining the MPS $ \vert O \rangle \rangle$.

Indeed, optimizing MPS overlaps can be done in an efficient manner, without resorting to parametric gradient computation ~\cite{schollwock2011density,white1992density,holtz2012alternating}, resulting in a considerable speed up with respect to gradient based methods. We employ this technique to swiftly find the best operator $|O\rangle \rangle$ that solves the classification task at hand. Notice that not only the data $|\tilde{\rho_i}\rangle\rangle$ lives in the low-bodyness subspace, but the MPS representing the evolved measurement operator is also constrained to the optimization within the low-bodyness subspace. This constraint reproduces the behavior of an average QCNN under random initialization, and proves to be sufficient to solve the task at hand. Hence, after optimization, the second phase returns an optimal MPS representing a low-weight measurement operator that is able to classify the training set. 
We can now proceed with the third and last step, which is optional with respect to the classification task, but which is needed to find the optimal parameters of the circuital ansatz we started with.
This step simply corresponds to compiling the ansatz $U(\theta)$ at hand, realized as a tensor network $\hat{U}(\theta)$ acting on the MPS representation of some final measurement operator $|M\rangle\rangle$, to be as close as possible to the found optimal $|O\rangle \rangle$. In the case considered in this manuscript, a 1-$d$ QCNN having logarithmic number of layers, the corresponding tensor network representation of the QCNN can never increase the bond dimension of the MPS it acts on by an exponential (in the number of qubits $n$) amount. Hence the full simulation and compilation of the QCNN is always efficient. In general, when dealing with hard-to-simulate ans\"{a}tze we can employ strategies akin to those described for the PP and PPS methods, basically interleaving the layers of the ansatz with projectors onto the low-body subspace. In any case, the compilation can be carried out optimizing the overlap $\langle\langle O|\hat{U}(\theta)|M\rangle\rangle$ between the optimal and parametrized Heisenberg evolved measurement operator. To this end, either numerical gradient methods or more precise tensor network techniques can be used \cite{kingma2015adam, schollwock2011density, holtz2012alternating}.

\section{Analysis of scaling and overfitting during training}

\label{ssec:scaling_numerics}

We choose the XXX model and the Haldane chain to showcase how the number of shadows required to perform successful training varies as a function of the system size. Moreover, in order to inspect as well the potential effect of overfitting, we modify the amount of samples employed in the training and testing datasets. In the main text, we used a dataset composed of 100 samples, using up to 75 of them as the training dataset for the XXX and Haldane models. Here, we enlarge the dataset to 500 samples, out of which 400 are employed in the training stage.

\begin{figure}[h]
\centering
    \begin{minipage}{0.49\columnwidth}
        \centering
        \includegraphics[width=\linewidth]{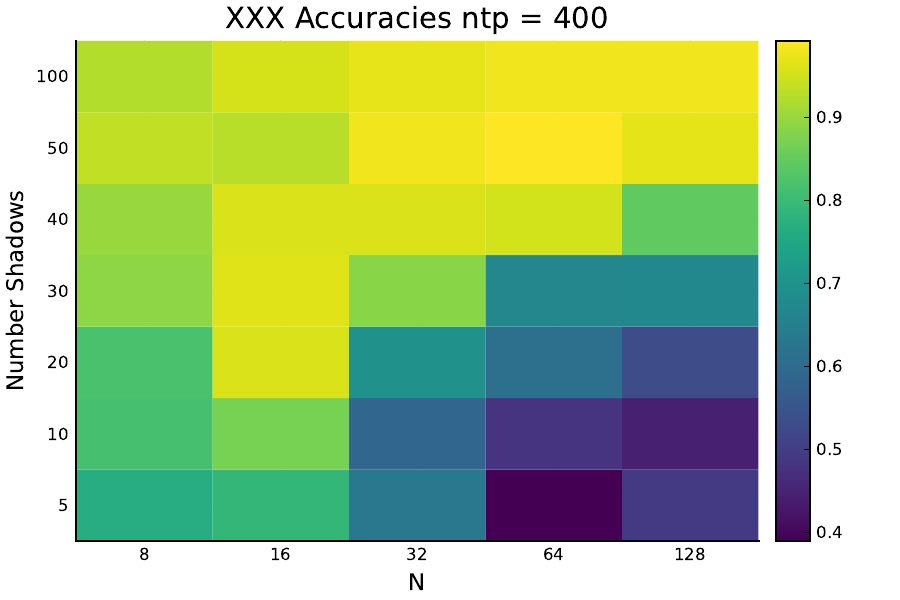}
        \caption{\textbf{Bond-Alternating XXX model.} Classification accuracy as a function of number of qubits and number of shadows.}
        \label{fig:XXX-diagram}
    \end{minipage}%
    \hfill
    \begin{minipage}{0.49\columnwidth}
        \centering
        \includegraphics[width=\linewidth]{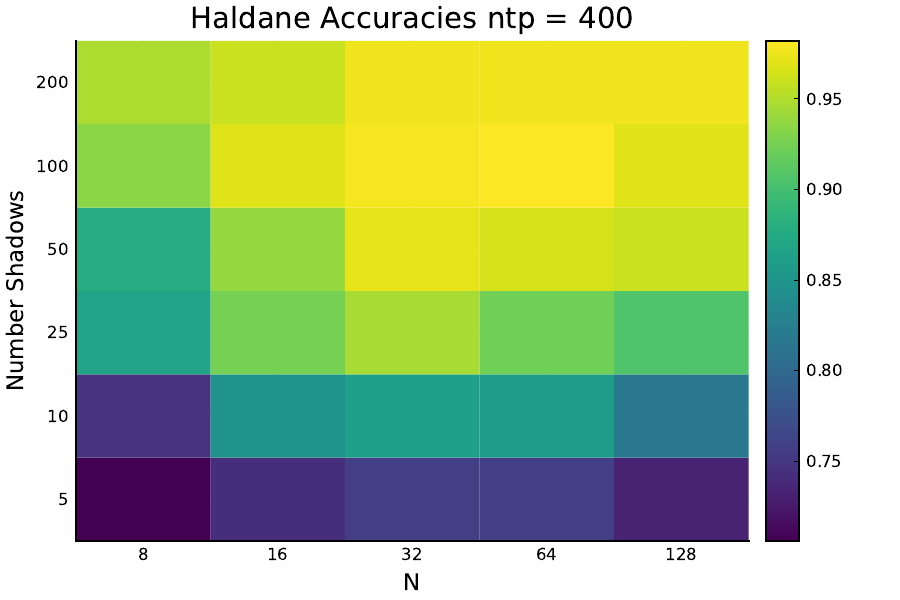}
        \caption{\textbf{Bond-Alternating Haldane model.} Classification accuracy as a function of number of qubits and number of shadows.}
        \label{fig:Haldane-diagram}
    \end{minipage}
\end{figure}

As we argued in the manuscript, which is further evidenced in these complementary results, the complexity of the classification does not substantially increase with system size. The number of shadows to perform successful classification in the XXX model increases sublinearly with system size, while there it seems to be independent of the system size in the case of the Haldane model.

One does not expect the intrinsic features of the data samples to change with system size, so that the kind of operators (in our case, amount and type of Pauli strings) employed to perform classification should not vary either. One can visualize this idea by considering a ferromagnetic material, where all spins are pointing at a given direction, and an antiferromagnetic one. A faithful measurement of 2-body operators (which can tell the difference in local magnetization) should be enough to tell phases apart. 

We would also like to point out that significantly more hyper-tuning can be conducted in the optimization process. In the case of the XXX model, where only 2-body operators were used for the classification, it could be the case that system size effects are stronger, thus explaining the sublinear increase in the number of shadows with system size. However, it is also possible that further hyper-tuning of the model (preselection of relevant Pauli strings, increase in number of iterations, etc.) lead to a classification process which barely depends on system size.

Lastly, we want to remark that no overfitting seems to be taking place in the optimization process, since we increase the number of training points and the number of shadows required to perform classification stays barely the same compared to the results shown in the main text with significantly lower number of training points. 

\section{Random Forest classification of the XXX Bond alternating model}

In this complementary analysis, we study the classification of the two ground states of the XXX Bond Alternating model using the random forest algorithm, a widely employed method in the machine learning literature. These results are aimed at showcasing that the dataset is locally easy, as we can perfectly classify the phases of matter by just considering single qubit Paulis and a few two-qubit ones.

\medskip

We generate 500 ground states from an $n=100$ qubit  XXX Bond Alternating model via DMRG(for J$_1$ = 1, 250 states belong to 0<J$_2$<1, and 250 states fall within 1<J$_2$<2). The features fed to the decision trees in the random forest are constructed as follows: the overlap of each ground state with all operators containing a single non-trivial (Pauli) operator on the string, plus 500 operators containing two non-trivial operators (with no preference for which positions the non-trivial operators occupy). We use 350 states for training (evenly split between phases and randomly distributed within each phase) and 150 states for testing. Since the number of features is moderate, we employ as many trees in the random forest as there are input features. Finally, we average the predicted labels over 200 independent trainings. The resulting classification is shown in Fig.~\ref{fig:random_forest}:

\begin{figure}
\centering
\includegraphics[width=0.5\linewidth]{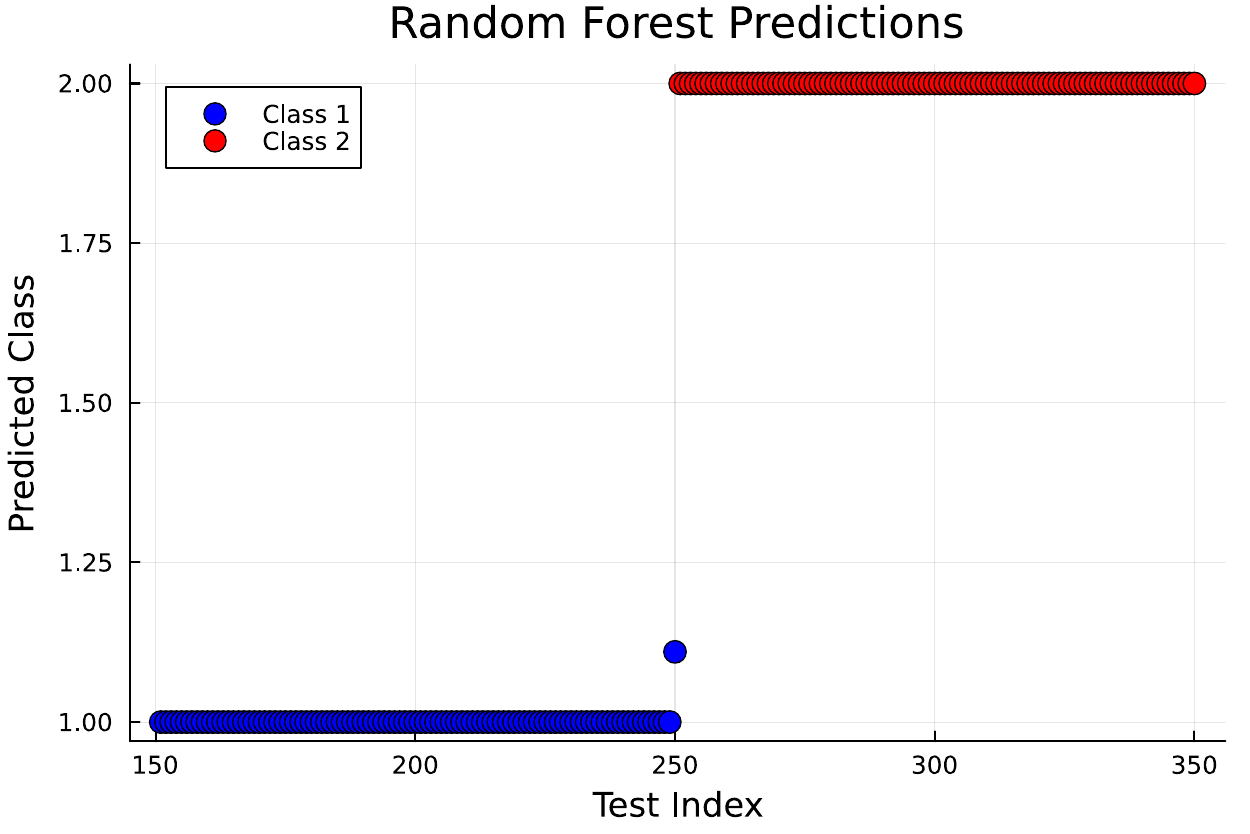}
\caption{Classification of ground states of the XXX Bond Alternating model with 
n=100 qubits using a random forest. Features include overlaps of the ground states with operators containing a single non-trivial (Pauli) operator (and identity elsewhere), plus 500 operators composed of two non-trivial operators.}
\label{fig:random_forest}
\end{figure}

\medskip

As clearly observed in Fig.~\ref{fig:random_forest}, the classification achieves 100\% accuracy. This indicates that the information contained in the module of a single non-trivial operator, complemented by some information from operators with two non-trivial operators, is sufficient to distinguish the two families. This result, entirely independent of the QCNN, demonstrates the intrinsic features of the datasets, making them easily classifiable.

\section{Measurement-based QCNN}\label{app:sec:meas-QCNN}

In this section we will study QCNNs where at each pooling layer, qubits are measured rather than traced out, and where the measurement outcomes are used to control unitaries applied to neighboring qubits~\cite{cong2019quantum}.

\subsection{Concentration properties and Heisenberg-evolved measurement operator}

\begin{figure}
\centering
\includegraphics[width=0.5\linewidth]{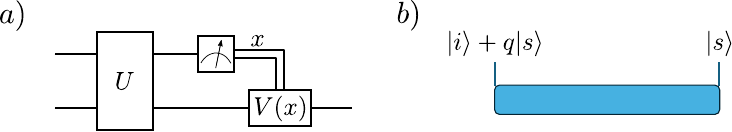}
\caption{(a) Unitary (from a convolutional layer) and the basic unit of the pooling layer: a measurement on a qubit, followed by a controlled unitary. (b) Adding the measurement in the pooling layer changes the input to the $P$ gate from $\ket{i}\ket{s}$ to $(\ket{i}+q\ket{s})\ket{s}$. }
\label{fig:qcnn_measuremntfig}
\end{figure}

First, let us define the two-qubit channel consisting of a random two qubit unitary $U$ followed by a measurement on the first qubit, and a controlled unitary $V(x)$ on the second qubit (see of Fig.~\ref{fig:qcnn_measuremntfig}(a)). As such, if the first qubit is measured on the state $x=0$ (or $1$), then we apply a unitary $V(0)$ (or $V(1)$) onto the second qubit. The action of this channel on a two-qubit state $\rho$ is given by
\begin{equation}
\mathcal{N}(\rho)=\sum_{x\in\{0,1\}}p(x) (\dya{x}\otimes V(x))U\rho U\ad  (\dya{x}\otimes V\ad(x))\,,
\end{equation}
where $p(x)=\Tr[(\dya{x}\otimes\id)\rho \dya{x}\otimes\id]$. Via vectorization, we can express the previous channel as
\begin{equation}
\widehat{\mathcal{N}}=\sum_{x\in\{0,1\}}p(x) (\dya{x}\otimes V(x))\otimes (\dya{x}\otimes V^*(x))\cdot\left(U\otimes U^* \right) \,.
\end{equation}

Assuming that $U$ is sampled according to the Haar measure over $\mathbb{U}(4)$, and taking the expectation value $\mathbb{E}[\cdot]=\int_{\mathbb{U}(4)}d\mu(U)[\cdot]$, we obtain the moment operator 
\begin{align}
\mathbb{E}[\widehat{\mathcal{N}}]&=\int_{\mathbb{U}(4)}d\mu(U)\sum_{x\in\{0,1\}}p(x) (\dya{x}\otimes V(x))\otimes (\dya{x}\otimes V^*(x))\cdot\left(U\otimes U^* \right) \nonumber\\
&=\sum_{x\in\{0,1\}}p(x) (\dya{x}\otimes \id)\otimes (\dya{x}\otimes \id)\cdot\int_{\mathbb{U}(4)}d\mu(U)\left(U\otimes U^* \right)\nonumber\\
&=\sum_{x\in\{0,1\}}p(x) (\dya{x}\otimes \id)\otimes (\dya{x}\otimes \id)\frac{| \id\otimes \id\rangle\rangle\langle\langle\id\otimes\id|}{4}\nonumber\,,
\end{align}
where in the second equality we have used the right- and left-invariance of the Haar measure and in the third we used Eq.~\eqref{eq:twirl-vectorized}. From the previous, we can see if we seek to estimate the expectation value of a traceless operator $O$ on the second qubit, then we need to compute  $\langle\langle\id\otimes\id|\id\otimes O\rangle\rangle=0$, which matches the expected value of zero as in the tracing out QCNN.

Then, the second moment operator, denoted as $\mathbb{E}[\widehat{\mathcal{N}}^{(2)}]$, can be found as~\cite{duschenes2024channel}
\begin{align}
\mathbb{E}[\widehat{\mathcal{N}}^{(2)}]
=&\left(\sum_{x,y\in\{0,1\}}p(x)p(y) (\dya{x}\otimes V(x))\otimes (\dya{y}\otimes V(y))\otimes (\dya{x}\otimes V^*(x))\otimes (\dya{y}\otimes  V^*(y))\right)\nonumber\\
\quad&\times\left(\int_{\mathbb{U}(4)}d\mu(U)U^{\otimes 2}\otimes (U^*)^{\otimes 2} \right)\,.\label{eq:meas-qcnn}
\end{align}
Note that here we cannot use the left- and right-invariance of the Haar measure to absorb the action of the unitaries $V(x)$ and $V(y)$. For instance, when $x\neq y$ we obtain terms of the form
\begin{equation}
    \int_{\mathbb{U}(4)}d\mu(U)(V(x)U)\otimes(V(y)U)\otimes (V^*(x)U^*)\otimes(V^*(y)U^*)\neq\int_{\mathbb{U}(4)} d\mu(U)U^{\otimes 2}\otimes (U^*)^{\otimes 2}\,.
\end{equation}
Instead, by using  Eq.~\eqref{eq:twirl-vectorized} we find (up to reordering of indexes)
\begin{align}
\int_{\mathbb{U}(4)}d\mu(U)U^{\otimes 2}\otimes (U^*)^{\otimes 2} =\frac{1}{15}\Big(&|\id_1\otimes \id_2\rangle\rangle\langle\langle \id_1\otimes \id_2|+|{\rm SWAP}_1\otimes {\rm SWAP}_2\rangle\rangle\langle\langle {\rm SWAP}_1\otimes {\rm SWAP}_2|\nonumber\\
&-\frac{1}{4}(|\id_1\otimes \id_2\rangle\rangle\langle\langle {\rm SWAP}_1\otimes {\rm SWAP}_2|+|{\rm SWAP}_1\otimes {\rm SWAP}_2\rangle\rangle\langle\langle\id_1\otimes \id_2|) \Big)\,,
\end{align}
where we recall that $\id_j$ and ${\rm SWAP}_j$ denote the identity and swap operators acting on the two copies of the $j$-th qubit. From the previous,  we see that we need to evaluate the terms
\begin{align}
    &(\dya{x}\otimes \dya{y})\otimes (V(x)\otimes V(y))\otimes (\dya{x}\otimes \dya{y})\otimes (V^*(x)\otimes V^*(y))|\id_1\otimes \id_2\rangle\rangle\nonumber\\
    &=(\dya{x}\otimes \dya{y})\otimes (\id\otimes \id)\otimes (\dya{x}\otimes \dya{y})\otimes (\id\otimes \id)|\id_1\otimes \id_2\rangle\rangle\nonumber\,,
\end{align}
and
\begin{align}
        &(\dya{x}\otimes \dya{y})\otimes (V(x)\otimes V(y))\otimes (\dya{x}\otimes \dya{y})\otimes (V^*(x)\otimes V^*(y))|{\rm SWAP}_1\otimes {\rm SWAP}_2\rangle\rangle\nonumber\\
         &=(\dya{x}\otimes \dya{y})\otimes (\id\otimes \id)\otimes (\dya{x}\otimes \dya{y})\otimes (\id\otimes \id)|{\rm SWAP}_1\otimes {\rm SWAP}_2\rangle\rangle\nonumber\,.
\end{align}
The previous equations show that when acting on the second Haar moment operator, the action of the $V(x)$ and $V(y)$ disappears thanks to the presence of the projectors on the first qubit. Hence, we need to evaluate the controlled unitary-independent term
\begin{align}
\mathbb{E}[\widehat{\mathcal{N}}^{(2)}]
=&\sum_{x,y\in\{0,1\}}p(x)p(y) (\dya{x}\otimes \id)\otimes (\dya{y}\otimes \id)\otimes (\dya{x}\otimes \id)\otimes (\dya{y}\otimes  \id)\left(\int_{\mathbb{U}(4)}d\mu(U)U^{\otimes 2}\otimes (U^*)^{\otimes 2} \right).\label{eq:cool-result}
\end{align}

At this point, we find it convenient to consider the previous term in the context of a Heisenberg-evolved measurement operator. For simplicity let us assume that the channel $\mathcal{N}$ is applied at the end of the circuit to the last two qubits prior to the measurement of the expectation value of a Pauli operator $O$ on the second qubit.  As seen in Fig.~\ref{fig:qcnn_measuremntfig}(b), this means that going into the second leg of the $P$ gate is an $\ket{s}$ operator coming from projecting $O$ into the $\ket{i}$ and $\ket{s}$ basis (where we recall that $\ket{i}=|\id\otimes\id\rangle\rangle$ and $\ket{s}=|{\rm SWAP}\rangle\rangle$). Then, the input to the first leg is the projection of $\dya{x}\otimes \dya{y}$ onto the $\ket{i}$ and $\ket{s}$ basis. One can readily find that 
\begin{equation}
|\dya{x}\otimes \dya{y}\rangle\rangle=|i\rangle +\delta_{x,y}\ket{s}\,,
\end{equation}
when adding the probabilities, we find that the input to the first leg is then
\begin{equation}
\sum_{x,y\in\{0,1\}}p(x)p(y) \left(|i\rangle +\delta_{x,y}\ket{s}\right)=|i\rangle+(p(0)^2+p(1)^2)\ket{s}=|i\rangle+q\ket{s}\,,
\end{equation}
where we  defined $q=p(0)^2+p(1)^2$ with $0\leq q\leq1$ (where the upper bound is saturated if and only if $p(x)=1$ for some $x$, indicating  that the measured qubit is a computational basis state, i.e., no entanglement between the two qubits).

Using similar arguments as the ones derived above, we now need to study how the input $\frac{2}{3}\ket{s}\bigotimes_{j=1}^{n-1}(|i\rangle+q_j\ket{s})$ spreads through the QCNN. Here, we defined $q_j=p_j(0)^2+p_j(1)^2$ and $p_j(x)$ is the probability of measuring qubit $j$ on state $x\in\{0,1\}$. At this point, we thus find it important to make several remarks. First, we can see that, adding measurements to the QCNN simply changes the measurement operator to be back-propagated as
\begin{equation}
\underbrace{\frac{2}{3}\ket{s}|i\rangle^{\otimes n-1}}_{\text{without measurements}}\quad \rightarrow \quad\underbrace{\frac{2}{3}\ket{s}\bigotimes_{j=1}^{n-1}(|i\rangle+q_j\ket{s})}_{\text{with measurements}}\,.\label{eq:op-to-backprop}
\end{equation}
Next, since the action of the convolutional layers remains the same with or without measurements (as per the right-most term of Eq.~\eqref{eq:cool-result}) the second moment of expectation values simply changes as 
\begin{equation}
\underbrace{\frac{2}{3}\langle\langle\rho^{\otimes 2}|{\mathcal{P}}\ket{s}|i\rangle^{\otimes n-1}}_{\text{without measurements}}\quad \rightarrow \quad\underbrace{\frac{2}{3} \langle\langle\rho^{\otimes 2}|{\mathcal{P}}\left(\ket{s}\bigotimes_{j=1}^{n-1}(|i\rangle+q_j\ket{s})\right)}_{\text{with measurements}}\,,
\end{equation}
where ${\mathcal{P}}$ denotes the moment operator composed of products of $P$ gates defined in Eq.~\eqref{eq:matrix-P}. In particular, it is clear that since the expectation value of $\langle\langle\rho^{\otimes 2}|$ with any operator in $\{\ket{i},\ket{s}\}$ is positive (as it simply corresponds to a purity in a reduced subsystem on  the qubits indicated by the $s$-indexes) then we find 
\begin{equation}
\frac{2}{3}\langle\langle\rho^{\otimes 2}|{\mathcal{P}}\ket{s}|i\rangle^{\otimes n-1}\leq \frac{2}{3} \langle\langle\rho^{\otimes 2}|{\mathcal{P}}\left(\ket{s}\bigotimes_{j=1}^{n-1}(|i\rangle+q_j\ket{s})\right)
\end{equation}
and hence,  if all gates in the convolutional layer form independent two designs
\begin{equation}
   {\rm Var}_{\vec{\theta}}\left[ \Tr[\Phi^{\rm tr}_{\vec{\theta}}(\rho_i)O]\right]\leq {\rm Var}_{\vec{\theta}}\left[ \Tr[\Phi^{\rm meas}_{\vec{\theta}}(\rho_i)O]\right]\,,\label{eq:concentration-comparison}
\end{equation}
where we defined the channels implementing the QCNN with tracing out as $\Phi^{\rm tr}_{\vec{\theta}}$, and with measurements as $\Phi^{\rm meas}_{\vec{\theta}}$. Equation~\eqref{eq:concentration-comparison} shows that the QCNN with measurements concentrates less than the QCNN without them, implying that adding measurements decreases the expressive power of the QCNN.  In particular, we note that the specific details of the convolutional layers are encoded into $\mathcal{P}$, and that Eq.~\eqref{eq:concentration-comparison} holds irrespective of how the trainable gates are arranged (i.e., one-dimensional QCNN, two-dimensional QCNN, etc). 

Another consequence of Eq.~\eqref{eq:op-to-backprop} is that when studying the bodyness in $\mathbb{E}_{\thv}\left[p^{(k)}_{{\Phi_{\thv}^{\rm meas}}\ad(O)} \right]$, only the local terms will dominate, implying that randomly-initialized measurement-based QCNNs can also only see local information in the input state. This is due to the fact that in ${\mathcal{P}}\ket{s}\bigotimes_{j=1}^{n-1}(|i\rangle+q_j\ket{s})=\mathcal{P}\left(\sum_{x\in\{0,1\}^{\otimes n-1}}Q^x\ket{s}\otimes_{j=1}^{n-1} \ket{s^{x_j}}\right)$ with $Q^x=q_1^{x_1}\times\cdots q_{n-1}^{x_{n-1}}$ and $\ket{s^{0}}=\ket{i}$, the coefficients decay exponentially with the number of $s$ in the Heisenberg-evolved operator. At this point, one may wonder about the case when $q_j=1$ $\forall j=1,\ldots,n-1$, as here the input to $\mathcal{P}$ is simply $\ket{s}^{\otimes n}$, which satisfies the property $\mathcal{P}\ket{s}^{\otimes n}=\ket{s}^{\otimes n}$ (as per the definition of the $P$ gates). In this pathological case, a global operator is able to move through the $P$-gates without becoming exponentially suppressed, and the QCNN could see the initial state globally.  However, it is easy to see that this case does not appear, as prior to the measurement, a Haar random two-qubit gate is applied. Thus, in average, the output of such gate is the maximally mixed state $\frac{\id\otimes \id}{4}$  from which one obtains $\mathbb{E}[p_j(0)]= \mathbb{E}[p_j(1)]=\frac{1}{2}$, and concomitantly $\mathbb{E}[q_j]=\frac{1}{2}$ for all measured qubits. Moreover, beyond the average case scenario, one only obtains $q_j=1$ for all $j$ if the input state to all measurements in an unentangled computational basis state, indicating that the QCNN is able to disentangle the input state via its simple convolutional layers. This implies an entanglement structure that can be readily reproduced by tensor networks, so that both the QCNN and the input data are classically simulable. In the next section, we discuss how tensor networks can still be used to simulate a measurement-based QCNN on more general input states.

\subsection{Classical simulability}

In the previous section, we have seen that randomly initialized measurement-based QCNNs will only ``see'' the local information in the initial state. Given that this information can be captured with Pauli classical shadows, we here explore whether classical shadows plus tensor networks enable for the efficient simulation of such QCNN architectures. Note that the fact that we use measurement information for feed-forward operations precludes the use of Heisenberg-evolution type techniques such as LOWESA or other Pauli propagation techniques. On the other hand, as measurements destroy entanglement in the system (pooling layers are in the class of local operations and classical communication, and thus, they cannot increase entanglement on average~\cite{horodecki2009quantum}), and as Pauli shadows lead to separable initial tensor product states, this combination actually enables efficient simulation vis standard MPS techniques.

In fact, we have explicitly performed the aforementioned simulation for a characteristic QCNN architecture, similar to the one employed for the numerical results of section \ref{section:numerical results}. Namely, the architecture consists of convolutional layers where two-qubit gates act on pairs of neighboring qubits in a brick-like fashion. Then, in the pooling layer, the qubits that will no longer be part of a convolutional operation are measured, and the measurement outcome controls a single qubit unitary on one of its nearest neighbors (e.g., if qubit $i$ is measured, then the controlled unitary is applied to qubit $i+1$). 

\begin{figure}
\centering
\includegraphics[width=1\linewidth]{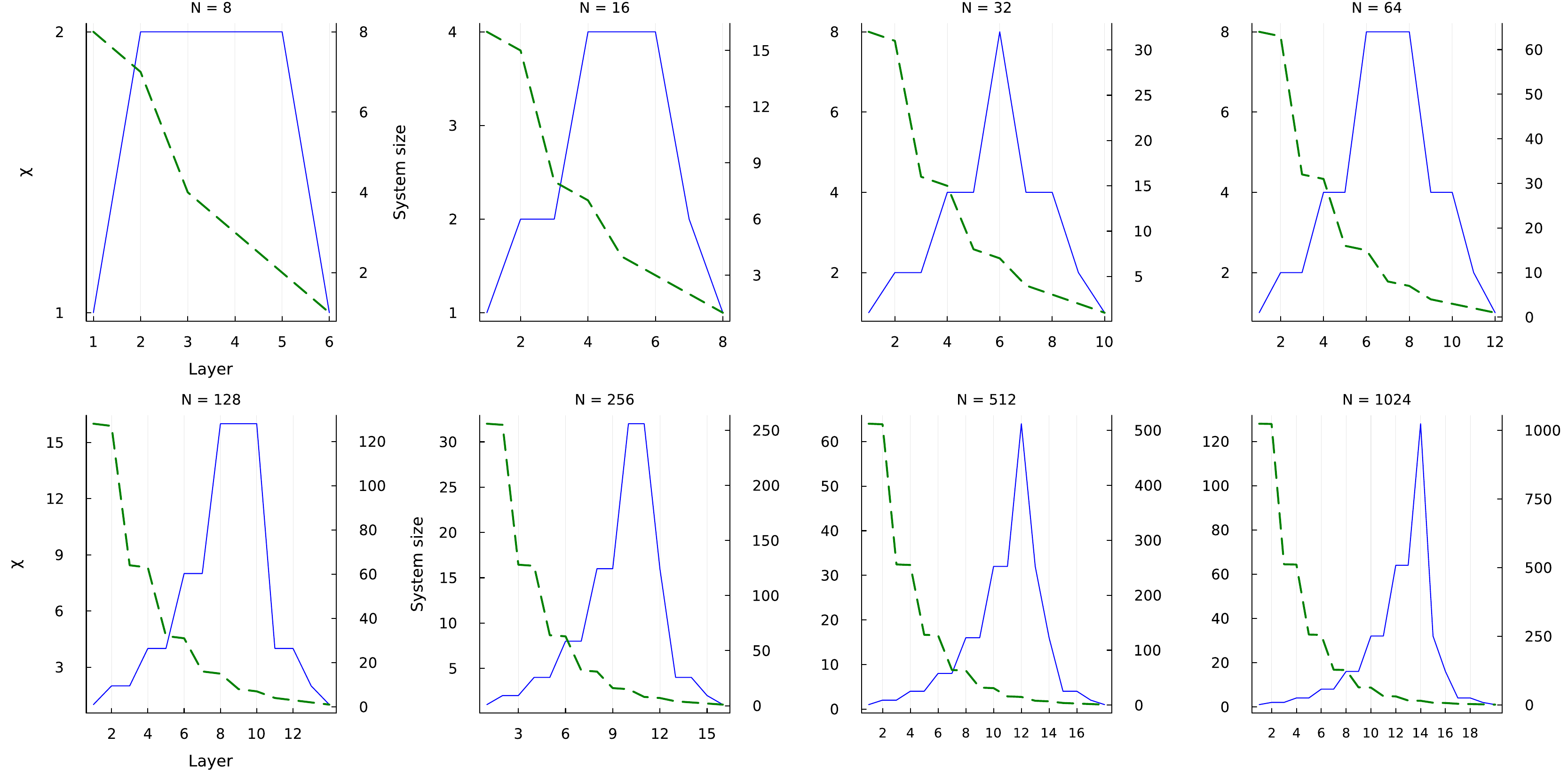}
\caption{Average bond dimension $\chi$ and systems size for a product state evolving through a typical $n$-qubit measurement-based QCNN. The horizontal axis depicts the layer number $l=0,\ldots\log(n)$, the solid blue curve the bond dimension, and the dashed green curve the system size of the quantum state at the $l$-th layer. Results are shown for $n=2^m$ for $m=3,4,\ldots,10$. All unitaries in the convolutional layer, as well as the control unitaries, were randomly sampled from $\mathbb{U}(4)$ and $\mathbb{U}(2)$, respectively.    }
\label{fig:chi_measurements}
\end{figure}

In Fig.~\ref{fig:chi_measurements} we show how the maximum bond dimension $\chi$ of the MPS scales for different system sizes as the initial state evolves through the QCNN. In this example, the initial state is a random product state, and the unitaries forming the convolutional layers, together with the unitaries from the controlled measurements, are randomly sampled from $\mathbb{U}(4)$ and $\mathbb{U}(2)$, respectively. From the plots, one can observe an initial increase in the bond dimension as the state is evolved throughout the convolutional layers, and initial pooling layers. Such growth, is eventually stopped by the fact that after each pooling layer half of the qubits are measured-out, effectively reducing the system size. After the crossover point, correlations start to decrease and are limited by the system size. At the end of the QCNN, a single-qubit state remains and the bond dimension is equal to its starting value of one.

We find that, at most, the bond dimension scales as $\chi \sim \tfrac{n}{8}$ . Since the MPS simulation cost scales as $\mathcal{O}(\chi^3)$, determined by the SVD performed after applying each gate, the total cost scales as $\mathcal{O}(n^3)$, i.e., polynomially in $n$. With the previous being said, it is worth mentioning that in practice even modest polynomial scaling can quickly become intractable. However, state-of-the-art methods allow MPS simulations on a laptop with bond dimension up to $\chi \sim 4096$, meaning that a laptop could simulate a QCNN with up to 32,768 qubits, far beyond the capabilities of current quantum computers. It is also worth noting that the previous simulation used random quantum gates in order to reproduce the worst-case scenario, whereas in practice one would expect smaller values of $\chi$ during a typical training process.

\end{document}